\documentclass{aa}  
\usepackage{graphicx}
\usepackage{txfonts}
\usepackage{lscape}
\usepackage{booktabs}
\usepackage{threeparttablex}
\usepackage{wasysym}
\usepackage{arydshln}
\usepackage{mathrsfs}
\usepackage{amssymb}
\usepackage[bookmarks=false,colorlinks=true,linkcolor=black,citecolor=cyan,filecolor=black,urlcolor=cyan]{hyperref}

\begin{document} 

    \title{Separating deterministic and stochastic gravitational wave signals in realistic pulsar timing array datasets}
\author{Irene Ferranti\inst{1,2}, Golam Shaifullah\inst{1,2,3}, Aurelien Chalumeau\inst{1,2,5} \& Alberto Sesana\inst{1,2,4}}

\institute{
    Dipartimento di Fisica ``G. Occhialini", Universit{\'a} degli Studi di Milano-Bicocca, Piazza della Scienza 3, I-20126 Milano, Italy\\
    \email{i.ferranti@campus.unimib.it}
    \and
    INFN, Sezione di Milano-Bicocca, Piazza della Scienza 3, I-20126 Milano, Italy
    \and
    INAF - Osservatorio Astronomico di Cagliari, via della Scienza 5, 09047 Selargius (CA), Italy
    \and
    INAF - Osservatorio Astronomico di Brera, via Brera 20, I-20121 Milano, Italy
    \and
    ASTRON, Netherlands Institute for Radio Astronomy, Oude Hoogeveensedijk 4, 7991 PD Dwingeloo, The Netherlands
    }
\titlerunning{Not anymore}
\authorrunning{Ferranti et al}

    
%
%

  \abstract{Recent observations by pulsar timing arrays (PTAs) suggest the presence of gravitational wave (GW) signals, potentially originating from supermassive black hole binaries (SMBHBs). These binaries can generate two kinds of signals: a stochastic gravitational wave background (GWB) or deterministic continuous gravitational waves (CGW). Being able to correctly recognize and separate them is crucial for accurate signal recovery and astrophysical interpretation. This paper aims at investigating the interaction between stochastic GWB and deterministic CGW signals within current analysis pipelines. We focus on understanding potential misinterpretations and biases in the parameter estimation when these signals are analysed separately or together. To this end, we performed several realistic simulations based on the European PTA 24.8yr dataset. We first injected either a GWB or a CGW into five datasets (three GWB realisations and two CGW realisations) with identical noise and analysed each signal type independently, and then analysed data sets containing both a stochastic GWB and a single resolvable CGW. We compared parameter estimation using different search models, including Earth term (ET) only or combined Earth and pulsar term (ET+PT) CGW templates, and correlated or uncorrelated power law GWB templates. We show that, when searched independently, the GWB and CGW signals can be misinterpreted as each other, and only a combined search is able to recover the true signal present. For datasets containing both a GWB and a CGW, failure to account for the latter biases the recovery of the GWB, but when performing a combined search, both GWB and CGW parameters can be recovered without any strong bias. Care must be taken with the method used to perform combined searches on these multi-component datasets, as the CGW PT can be misinterpreted as a common uncorrelated red noise. However, this can be avoided by direct searches for a correlated GWB plus a CGW (ET+PT). Our study underscores the importance of combined searches to ensure unbiased recovery of GWB parameters in the presence of strong CGWs. This is crucial to interpret the signal recently found in PTA data, and is a first step towards a robust framework for disentangling stochastic and deterministic GW components in future, more sensitive datasets.}
    

   \maketitle
%

\section{Introduction}

Pulsar Timing Array (PTA) collaborations search for gravitational waves (GWs) with frequencies in the nanoHertz (nHz) band ($10^{-9}-10^{-7}$Hz). The primary target of PTA experiments is a stochastic gravitational wave background (GWB) generated by the incoherent superposition of the GWs emitted by all the super-massive black hole binaries (SMBHBs) that populate our universe \citep{1995ApJ...446..543R,Jaffe2003,2003ApJ...590..691W,2008MNRAS.390..192S}. The signal manifests itself in the data as a red noise process common to all the pulsars in the array, with a specific angular correlation, first derived by \cite{1983_HD}. Recently, the PTA community found evidence for a signal with both these characteristics: the collaborations reporting this finding were the European PTA collaboration along with the Indian PTA collaboration \citep[EPTA, InPTA][]{EPTA}, the North American Nanohertz Observatory for Gravitational Waves collaboration \citep[NANOGrav,][]{NANOGrav}, the Parkes PTA collaboration \citep[PPTA,][]{PPTA} and the Chinese PTA collaboration \citep[CPTA,][]{CPTA}. 

While the Hellings and Downs (HD hereinafter) correlation is distinctive of a GW origin of the signal, separating it from possible correlated noise of different nature \citep{2016MNRAS.455.4339T}, its source cannot yet be confidently established. In fact, besides SMBHBs, a nHz cosmic GWB can also be produced by a number of physical processes occurring in the early universe, such as cosmic strings networks \citep{Damour_2000}, primordial curvature perturbations \citep{1967PThPh..37..831T}, or QCD phase transitions in the early universe \citep{1992PhRvL..69.2026K}, all of which can provide a viable explanation of the observed signal \citep{2023ApJ...951L..11A,2024A&A...685A..94E}.
Moreover, the HD correlation is not unique of stochastic GWBs; as shown in \cite{Cornish_2013} and in \cite{2023PhRvD.107d3018A}, the signal generated by a single SMBHB and observed by an array of pulsars results in an angular correlation with the HD expectation value, albeit with a somewhat different variance. Note that the quality of the current data does not allow a description of the signal variance and  therefore, the possibility that a particularly loud source -- a massive and nearby SMBHB -- can be mistaken by a GWB is real. By loud source, we mean a source that can be singled out from the noise and, if present, from the GWB by means of a deterministic, template based search. Since the SMBHBs producing such signals are far from mergers and persistent in the data, their signal is generally referred to as continuous gravitational wave \citep[CGW,][]{Babak_2012}.

CGWs searches have been performed on the latest data release of EPTA \citep[DR2new,][]{CGW_EPTA} and on the NANOGrav 15yr dataset \citep{2023ApJ...951L..50A}. They both found evidence in favour of the presence of a CGW with a 2.5-3$\sigma$ significance, compared to a model where only pulsar noise is present in the data. This value is comparable to the evidence in favour of the presence of a GWB, which has a 3-to-4$\sigma$ significance). Some simulations carried out in \cite{CGW_EPTA} already showed how a CGW can be misinterpreted as a GWB and, vice versa, a GWB can be misinterpreted as a CGW. In both the EPTA and NANOGrav analyses, when a GWB is also searched for, evidence for a CGW significantly decreases, indicating that its presence is at the very least not necessary to explain the observations. However, the degeneracy between the pulsar response to the two signals does not allow to conclusively rule out the presence of a CGW. 

The presence of these resolvable sources in the PTA data on the one hand poses a number of interesting challenges, on the other hand opens the door to invaluable opportunities. In fact, while the analysis pipelines currently used to search for a GWB rely on the assumption that the background is a Gaussian, isotropic, and stationary process characterised by a power-law Fourier spectrum \citep{2015MNRAS.453.2576L,2016ApJ...821...13A}, the presence of a CGW can result in an excess of power at specific frequencies, which produces some level of anisotropy and non Gaussianity \citep[e.g.,][]{2024ApJ...965..164G,2024arXiv240414508S}. Thus, the standard analysis may depend on inapplicable assumptions and yield biases on the estimated GWB parameters \citep{2023ApJ...959....9B,2024A&A...683A.201V}. This is a recognised issue within the PTA community and analysis techniques are being developed to go beyond the power law, isotropic Gaussian assumption \citep[e.g.,][]{2013PhRvD..88h4001T,2014PhRvD..90h2001G,2020PhRvD.102h4039T,2020PhRvD.102l2005A,2021ApJ...915...97X,2022ApJ...941..119B} and to pull out of that data GWBs and CGWs simultaneously \citep{2020CQGra..37m5011B,2024arXiv240711135F}.

Within this context, it is of paramount importance to study the interplay of CGWs and GWB when both are present in the data and what challenges arise in the interpretation of the analysis results. From an exquisitely scientific perspective, conversely, the prospects of detecting CGWs is particularly appealing. In fact due to their deterministic nature, PTAs are able to estimate the source parameters, in particular the sky location via triangulation \citep{Babak_2012,2012ApJ...756..175E} and possibly the distance to the source \citep[especially if the pulsar term of the induced timing delay can be measured,][]{Ellis_2013}. The detection of one or more resolvable sources \citep[which is expected in the SKA PTA era,][]{2024arXiv240712078T} would be very precious as it would allow to track the orbital evolution of the binary and thus obtain direct information on the Massive Black Holes (MBHs) coupling to the environment \citep[e.g.][]{2011MNRAS.411.1467K,2013CQGra..30v4014S,2014MNRAS.442...56R}. This information cannot be well constrained by a measure of the GWB power spectral density, because it is strongly degenerate with the MBH mass and eccentricity distribution. In addition, CGWs offer the exciting opportunity to search for electromagnetic counterparts of the GW signal, opening the prospects of low frequency GW multimessenger astronomy \citep{2012MNRAS.420..860S,2013CQGra..30v4013B,2019BAAS...51c.490K,2021ApJ...921..178L,2022MNRAS.510.5929C}. Although PTAs can realistically localise the CGW source only within several deg$^2$ \citep{2010PhRvD..81j4008S,2018MNRAS.477.5447G} they must be originated by particularly massive and/or nearby SMBHBs, making the search of their galaxy host viable \citep{2019MNRAS.485..248G,2024arXiv240604409P}. Identifying the host and detecting accretion activity onto the binary will shed light about the physics governing the formation and (thermo)dynamical evolution of SMBHBs and the accretion discs surrounding them. 

In this paper, we carry out the first detailed investigation of the interplay between stochastic GWBs and deterministic CGWs in realistic PTA datasets, assessing the performance of currently used analysis pipelines when both signals are present in the data. We do so by building state of the art mock datasets capturing the complexity of real data, and by injecting them with as realistic as possible GW signals, mimicking the one found in the recent EPTA analyses \citep{EPTA, 2023, CGW_EPTA}. Realistic GWBs are injected as the incoherent sum of individual waves consistent with astrophysical observations \citep{Rosado_2015}, and CGW are modelled according to the candidate identified in \cite{CGW_EPTA}. Datasets are then analysed with the standard models implemented in the \texttt{ENTERPRISE} \citep[Enhanced Numerical Toolbox Enabling a Robust PulsaR Inference SuitE,][]{enterprise, enterprise_extension} analysis suite, which models the GWB as a common red process characterized by a single power law spectrum, while we also test CGW models including only the Earth term or the full Earth plus pulsar components. 

The paper is organized as follows. In Section \ref{sec:methods} we describe in detail the simulations: the properties of the adopted datasets, the injected signals and the model used for the recovery. In Section \ref{sec:res} we present the main results of our analysis. Using the methods described in the previous section, we carried out several investigations. We first assess the importance of including the pulsar term in the CGW model (Section~\ref{sec:etvspt}). We then study the degeneracy of signal recovery on datasets containing only a GWB or a CGW (Section \ref{GWBvsCGW}), showing that datasets injected only with a CGW can produce evidence of a GWB (Section \ref{CGWasGWB}) and vice versa (Section \ref{GWBasCGW}), and that a joint search is able to pin down the correct nature of the signal (Section \ref{combined}). The performance of a combined analysis of a GWB and a CGW on datasets containing both signals is then evaluated in Section \ref{combined_searchanddata} and we discuss our main findings in Section \ref{sec:discussion}.\\ Throughout the paper we assume a $\Lambda$ cold dark matter cosmology with parameters $(H_0, \Omega_m, \Omega_{\Lambda})=(70{\rm km s^{-1}Mpc^{-1}}, 0.3, 0.7)$.


\section{Methods}\label{sec:methods}
In this section, we describe both the simulations structure and the models and methods used in the analysis.

\subsection{Simulations}

\begin{figure*}
   \centering
   \includegraphics[width=1\textwidth]{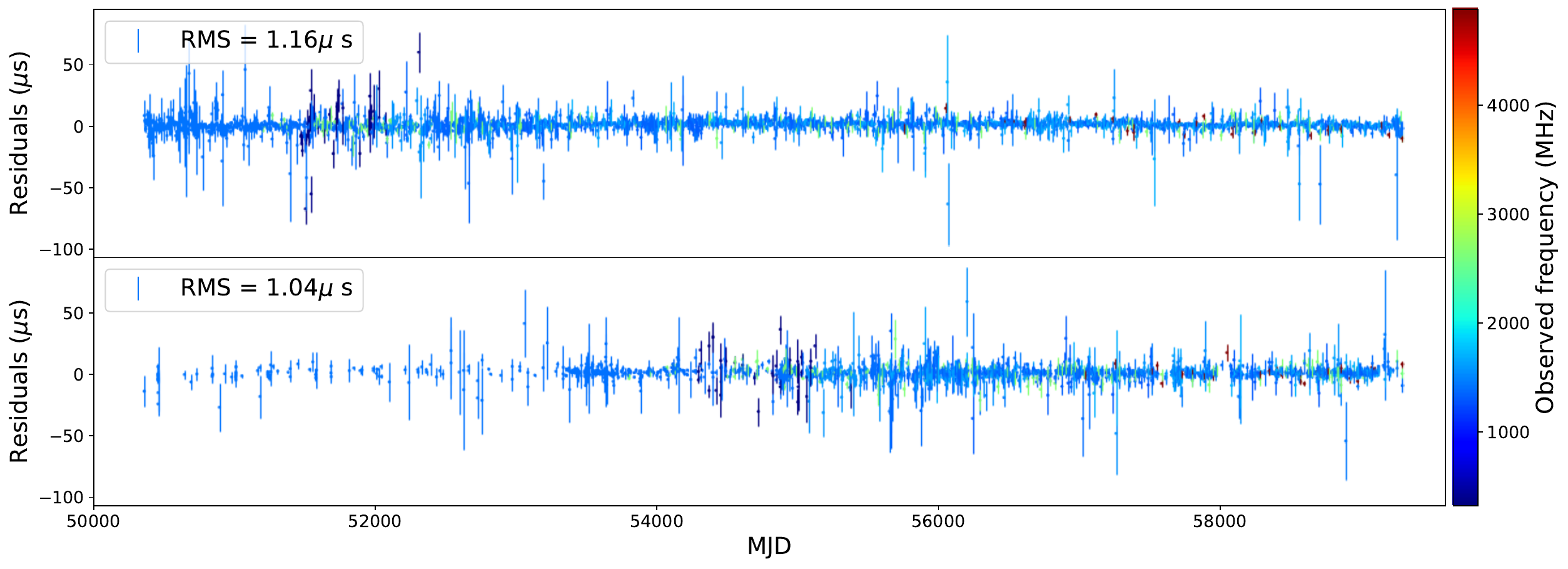}
   \caption{Pulsar PSR J1022+1001. \textit{Top:} simulated data. \textit{Bottom:} real data from EPTA DR2full. The properties that are exactly reproduced are the time span, the number of observations, the frequency coverage and the noise levels. The main difference between the two datasets is in the distribution of the ToAs: while in our simulations ToAs are unevenly sampled, they are still uniformly distributed over the observational timespan, meaning that sparse observations and long gaps are absent (as can be clearly seen by visual comparison of the two panels).
   }
   \label{fig:J1022}%
\end{figure*}

Our simulations are based on the second data release of EPTA \cite{2023}. In particular, we have reproduced the DR2full dataset: 25 pulsars with a maximum observation time of 24.8 yr. The main properties of the real time of arrivals (ToAs) of each pulsar are reproduced: the observation time, the number of ToAs, the timing model (TM), the multiband measures -- the frequency coverage is identical to that of the EPTA DR2full pulsars -- and the noise levels. The uneven sampling of the ToAs is also reproduced, but large gaps, which are common in the real data streams, are not present in the simulations (see Fig.~\ref{fig:J1022} for a comparison between real and simulated residuals).
All the properties of the arrival times were injected in the dataset using the \texttt{libstempo} package \citep{2020ascl.soft02017V}. The same package is used to inject the three components of the noise budget: white noise (WN), achromatic -- independent of the observation radio frequency --  red noise (RN) and chromatic red noise (due to dispersion measure variations, and customarily referred to as DM). 

To model the white noise, the ToAs are sampled from a Gaussian distribution centred at the value predicted by the TM and with a width given by:
\begin{equation}\label{wn}
    \sigma_i = \sqrt{EFAC_i^2\sigma_{\rm ToA}^2 + EQUAD_i^2},
\end{equation}
where $\sigma_{\rm ToA}$ is the errorbar associated to each ToA according to the template-fitting errors obtained in EPTA DR2, the EFAC takes into account for the ToA measurement errors and the EQUAD accounts for any putative extra source of white noise. The two parameters EFAC and EQUAD are specific for each observing backend. In the simulations, each pulsar's backend, as well as the ToAs' uncertainties, are identical those of DR2full; therefore, each pulsar has ToAs taken from $\sim$12 different backends, which means 12 different couples of values for EFAC and EQUAD. The values used for EFAC and EQUAD are the maximum likelihood values obtained by the single pulsar noise analysis of EPTA DR2full \citep{2023}.\\
In addition to the white noise, achromatic and chromatic red noise contribute to the single-pulsar noise budget. These are time-correlated noise components modelled as a stationary Gaussian process with a power-law power spectral density (PSD) of the form
\begin{equation}\label{rn_model}
    P(f; A, \gamma) = \frac{A^2}{12\pi^2}\biggl(\frac{f}{{\rm yr}^{-1}}\biggr)^{-\gamma}{\rm yr}^3.
\end{equation}
For achromatic red noise, in each ToA, the delay induced in the residuals is proportional to the red noise PSD, 
\begin{equation}\label{eg:RN_res}
    \Delta_{\rm RN}\propto \sqrt{P(f; A_{\rm RN}, \gamma_{\rm RN})}.
\end{equation}
This noise component accounts for the long-term variability of the pulsar spin along with other sources, such as unmodelled objects orbiting the pulsar. For chromatic red noise, on the other hand, the induced delay also depends on the observing radio frequency because it is due to the interaction of the radio signal with the ionised interstellar medium (IISM), the interplanetary medium in the solar system, and the Earth's ionosphere. This effect is taken into account during the observations and inside the timing model which considers its value at a reference epoch together
with its first and second derivatives. However, the inhomogeneous and turbulent nature of the IISM also induces stochastic variations in the DM value, which are modelled as chromatic red noise. The residual DM delay is modelled as 
\begin{equation}\label{DM_freq}
    \Delta_{\rm DM} \propto \sqrt{P(f; A_{\rm DM}, \gamma_{\rm DM})}\times\nu^{-2}
\end{equation}
where $\nu$ is the observing radio frequency. The values of $A_{\rm RN}$, $\gamma_{\rm RN}$, $A_{\rm DM}$ and $\gamma_{\rm DM}$ used in the simulations are  reported in Table \ref{noise_model}, together with the number of Fourier components used in the modelling of these processes. These values -- amplitude, slope and number of Fourier components -- correspond to maximum likelihood values obtained from the single pulsar noise analysis of EPTA DR2full. An example of simulated data compared to the real EPTADR2 dataset is shown in Fig. \ref{fig:J1022}.
\\
\begin{table*}
\centering
      \begin{tabular}{|c|c|c|c|c|c|c|c|c|c|}
      \hline
        Pulsar	&	T$_{\rm obs}$ [yr]	&	RMS$_{\rm ToAs}$ [$\mu s$]	&	Noise model	&	log$_{10}$A$_{\rm RN}$	&	$\gamma_{\rm RN}$	&	n$_{\rm RN}$	&	log$_{10}$A$_{\rm DM}$	&	$\gamma_{\rm DM}$	&	n$_{\rm DM}$	\\\hline\hline
J0030+0451	&	22	&	2.47	&	RN	&	-15.18	&	5.71	&	10	&	-	&	-	&	-	\\
J0613-0200	&	22.9	&	3.4	&	RN + DM	&	-14.87	&	5.05	&	10	&	-13.51	&	2.13	&	144	\\
J0751+1807	&	24.2	&	2.11	&	DM	&	-	&	-	&	-	&	-13.52	&	3.17	&	60	\\
J0900-3144	&	13.6	&	3.08	&	RN + DM	&	-12.76	&	1.09	&	138	&	-13.48	&	3.43	&	151	\\
J1012+5307	&	23.7	&	1.43	&	RN + DM	&	-13.05	&	1.2	&	149	&	-13.6	&	1.62	&	45	\\
J1022+1001	&	24.5	&	1.16	&	RN + DM	&	-13.77	&	3.05	&	30	&	-13.22	&	0.02	&	100	\\
J1024-0719	&	23.1	&	1.41	&	DM	&	-	&	-	&	-	&	-13.48	&	2.31	&	34	\\
J1455-3330	&	15.7	&	2.74	&	RN	&	-13.18	&	1.67	&	49	&	-	&	-	&	-	\\
J1600-3053	&	14.3	&	2.18	&	DM	&	-	&	-	&	-	&	-13.82	&	3.89	&	26	\\
J1640+2224	&	24.4	&	1.27	&	DM	&	-	&	-	&	-	&	-13.44	&	0.41	&	146	\\
J1713+0747	&	24.5	&	1.04	&	RN + DM	&	-13.95	&	2.76	&	11	&	-13.5	&	1.63	&	148	\\
J1730-2304	&	16.1	&	1.34	&	DM	&	-	&	-	&	-	&	-13.27	&	2.49	&	10	\\
J1738+0333	&	14.1	&	3.2	&	RN	&	-12.93	&	1.89	&	11	&	-	&	-	&	-	\\
J1744-1134	&	24	&	1.25	&	RN + DM	&	-16.16	&	6.44	&	9	&	-13.41	&	0.93	&	151	\\
J1751-2857	&	14.7	&	3.44	&	DM	&	-	&	-	&	-	&	-12.76	&	2.3	&	41	\\
J1801-1417	&	13.7	&	4.88	&	DM	&	-	&	-	&	-	&	-12.3	&	1.33	&	14	\\
J1804-2717	&	14.7	&	2.85	&	DM	&	-	&	-	&	-	&	-12.8	&	0.66	&	38	\\
J1843-1113	&	16.8	&	4.18	&	DM	&	-	&	-	&	-	&	-12.67	&	1.98	&	73	\\
J1857+0943	&	24.1	&	2.44	&	DM	&	-	&	-	&	-	&	-13.41	&	2.7	&	10	\\
J1909-3744	&	15.7	&	1.17	&	RN + DM	&	-15.14	&	5.13	&	20	&	-13.61	&	1.45	&	151	\\
J1910+1256	&	15.2	&	3.32	&	DM	&	-	&	-	&	-	&	-13.28	&	2.84	&	10	\\
J1911+1347	&	14.2	&	2.12	&	DM	&	-	&	-	&	-	&	-13.64	&	3.18	&	10	\\
J1918-0642	&	19.7	&	2.32	&	DM	&	-	&	-	&	-	&	-13.94	&	3.81	&	138	\\
J2124-3358	&	16	&	2.45	&	DM	&	-	&	-	&	-	&	-13.07	&	1.03	&	41	\\
J2322+2057	&	14.7	&	4.44	&	-	&	-	&	-	&	-	&	-	&	-	&	-	\\
\hline
      \end{tabular}
      \vspace{0.2cm}
\caption{For each pulsar in the array, the following properties are listed: name of the pulsar, observation time span, RMS of the timing residual, contributions to the noise model found through model selection in the single pulsar noise analysis of EPTA DR2full and the maximum likelihood values of the slope, the amplitude and the number of Fourier bins of RN and DM as obtained from the  noise analysis of EPTA DR2full \citep{2023}. These values are used in the noise injection.}
\label{noise_model}
\end{table*}

\subsubsection{Gravitational wave signal injection}

Having specified the properties of the adopted PTA, we now turn to the description of the injected GW signals. The signal produced by a SMBHB is composed by two terms, customarily referred to as Earth term (ET) and pulsar term (PT), which are respectively associated to the GW strain hitting the Earth at the ToA of the considered pulse and the strain hitting the pulsar at the time of emission of the same pulse.

Throughout the paper we will always assume circular SMBHBs, evolving under the effect of GW backreaction only (i.e. we neglect any possible coupling with the local stellar or gaseous environment). Under these assumptions, the delays induced in the ToAs at an epoch $t$ are fully defined by 8 parameters: the chirp mass $\mathcal{M}$\footnote{\footnotesize{Note that in the quadrupolar approximation employed here for the injected waveform, the signal does not depend on the individual binary masses.}}, the observed frequency $f_{gw}$, the redshift $z$, the inclination angle $\imath$, the sky location $(\theta, \phi)$ -- respectively polar and azimuthal angle -- , the initial phase $\phi_0$ and the polarization angle $\psi$. Following \cite{Ellis_2013}, we model them as:
\begin{equation} \label{eq:SMBHB_delays}
s(t, \hat{\Omega}) = F^+(\hat{\Omega})\Delta s_+(t) +F^{\times}(\hat{\Omega})\Delta s_{\times}(t),
\end{equation}
where 
\begin{equation}
\Delta s_A(t) = s_A(t_p) - s_A(t_e),
\end{equation}
the functions $F^A(\hat{\Omega})$ -- with $A$ indexing the polarization and $\hat{\Omega}$ being the unit vector pointing from the GW source to the Solar System Barycenter ˆ
(SSB) -- are the antenna pattern functions that depend on the relative position of the pulsar and the source \citep[see][for full details]{Ellis_2013}. $s_A(t_p)$ and $s_A(t_e)$ are the PT and ET respectively. They both take the form

\begin{equation}\begin{split}
    &s_{+}(t) = \frac{\mathcal{M}^{5/3}}{d_L\omega(t)^{1/3}}\biggl(-\text{sin}(2(\Phi(t)-\Phi_0))(1+\text{cos}^2\imath)\text{cos}(2\psi) \\ &- 2\text{cos}(2(\Phi(t)-\Phi_0))\text{cos}\imath\, \text{sin}(2\psi)\biggr)\\
    &s_{\times}(t) = \frac{\mathcal{M}^{5/3}}{d_L\omega(t)^{1/3}}\biggl(-\text{sin}(2(\Phi(t)-\Phi_0))(1+\text{cos}^2\imath)\text{sin}(2\psi) \\ &+2\text{cos}(2(\Phi(t)-\Phi_0))\text{cos}\imath\, \text{cos}(2\psi)\biggr),
\end{split}\end{equation}
where 
\begin{equation}
    \Phi(t) = \Phi_0 + \frac{1}{32\mathcal{M}^{5/3}}\biggl(\omega_0^{-5/3}-\omega(t)^{-5/3}\biggr),
\end{equation}
and 
\begin{equation}\label{eq:omega}
    \omega(t) = \biggl(\omega_0^{-8/3}-\frac{256}{5}\mathcal{M}^{5/3}t\biggr)^{-3/8}.
\end{equation}
The only difference between PT and ET is that while the latter is the delay induced at the time of observation on Earth $t_e$, the former has to be evaluated at the time of emission of the radio pulse at the pulsar $t_p = t_e - L(1 + \hat{\Omega}\cdot\hat{p})$, where $L$ is the pulsar distance and $\hat{p}$ is the pulsar position in the sky. We inject in our data two different GW components: a stochastic GWB and a resolvable CGW. Since the evolution of the signal over the observation time is small, the phase approximation is used both in the injection and in the recovery: $\omega_{\rm Earth} = \omega_0$ at each epoch $t$.

{\it Stochastic GWB.} The GWB is injected as a realistic superposition of GWs coming from a cosmic population of SMBHBs. The injection pipeline starts from a catalog of SMBHBs, computes the delays induced by each binary in each pulsar ToA according to Eq.~\eqref{eq:SMBHB_delays} and then sums all the delays in the time domain. The distribution of the sources' $\mathcal{M}$, $f_{gw}$, $z$ and $\imath$ determines the observed power spectrum generated by the SMBHBs population: the binned total characteristic strain, where the frequency bin width is $\Delta f = 1/T$ with $T$ the time span of the pulsar array, can be evaluated as \citep{Rosado_2015}
\begin{equation}\label{eq:hc_bin}
    h_{c}^{2}(f_i) = \sum_{j\in \Delta f_i}\frac{h_j^2(f_r)f_r}{\Delta f_i},
\end{equation}
where $f_i$ is the central frequency of the bin $\Delta f_i$, and the sum runs over all the binaries for which the observed frequency $f=f_r/(1+z)\in \Delta f_i$, where $f_r$ is the GW frequency in the binary rest frame. The strain of each GW signal is given by 
\begin{equation}
    h(f_r) = 2\sqrt{\frac{1}{2}\bigl(a(\imath)^2 + b(\imath)^2\bigr)}\frac{(G\mathcal{M})^{5/3}(\pi f_r)^{2/3}}{c^4d(z)},
\end{equation} 
where $a(\imath) = 1+\text{cos}^2\imath$, $b(\imath) = -2\text{cos}\imath$ and $d(z)$ is the comoving distance of the binary, which is computed as 
\begin{equation}
    d(z) = \frac{c}{H_0}\int_{0}^{z}\frac{1}{\sqrt{\Omega_m (1+z)^3 + \Omega_{\Lambda}}}dz.
\end{equation}
For the population of circular GW driven SMBHBs, assumed here, the characteristic strain given by Eq.~\eqref{eq:hc_bin} can be approximated as a power law. It is therefore customary to describe the signal as
\begin{equation}\label{eq:hc_phinney}
    h_{c}(f) = A_{\rm GW}\left(\frac{f}{{\rm yr}^{-1}} \right)^{\alpha},
\end{equation}
where $\alpha=(3-\gamma)/2=-2/3$ being $\gamma$ the PSD induced in the timing residual (cf. Eq.~\ref{rn_model}).

The power spectrum $h_{c}^{2}(f_i)$ thus depends on the number of emitting sources per unit redshift, mass and frequency $d^3N/(dzd\mathcal{M}df_r)$. A list of $\sim$ 120k binaries is constructed by extracting $\mathcal{M}$, $f_{gw}$, $z$ as Monte Carlo samples of a numerical distribution $d^3N/(dzd\mathcal{M}df_r)$, obtained from the observation-based models described in \cite{Sesana_2013} and \cite{Rosado_2015}. In those papers, the authors construct hundreds of thousands of SMBHB populations in agreement with current constraints on the galaxy merger rate, the relation between SMBHs and their hosts, the efficiency of SMBH coalescence and the accretion processes following galaxy mergers. From this set of models, we selected 100 populations producing a GWB amplitude $A \sim 2.4\times10^{-15}$, consistent with the signal observed by EPTA \citep{EPTA}. The other parameters of the binaries are then randomly sampled such that sources are isotropically distributed in the sky -- $\text{cos}\,\theta \in \text{Uniform}(-1, 1)$ and $\phi \in \text{Uniform}(0, 2\pi)$ --, their orbital angular momenta are uniformly distributed on the unit sphere -- $\text{cos}\,\iota \in \text{Uniform}(-1, 1)$ --, the phase and the polarization angle are random -- $\phi_0 \in \text{Uniform}(0, 2\pi)$, $\psi \in \text{Uniform}(0, \pi)$. 

{\it Resolvable CGW.} In some of the dataset, an individually resolvable CGW is injected. The waveform is again injected in each pulsar in the time domain residual according to Eq.~\eqref{eq:SMBHB_delays}. The only difference, compared to the SMBHBs contributing to the unresolved GWB is that the amplitude of the signal is large enough that the source can be identified above the pulsar noise and the unresolved GWB by a signal search employing a deterministic filter to match the signal (see next Section). We stress that we always inject both the ET and the PT for every SMBHB, regardless of whether it provides an unresolvable contribution to the GWB or it is a resolvable CGW.

\begin{figure*}
    \centering
    \includegraphics[width=1\textwidth]{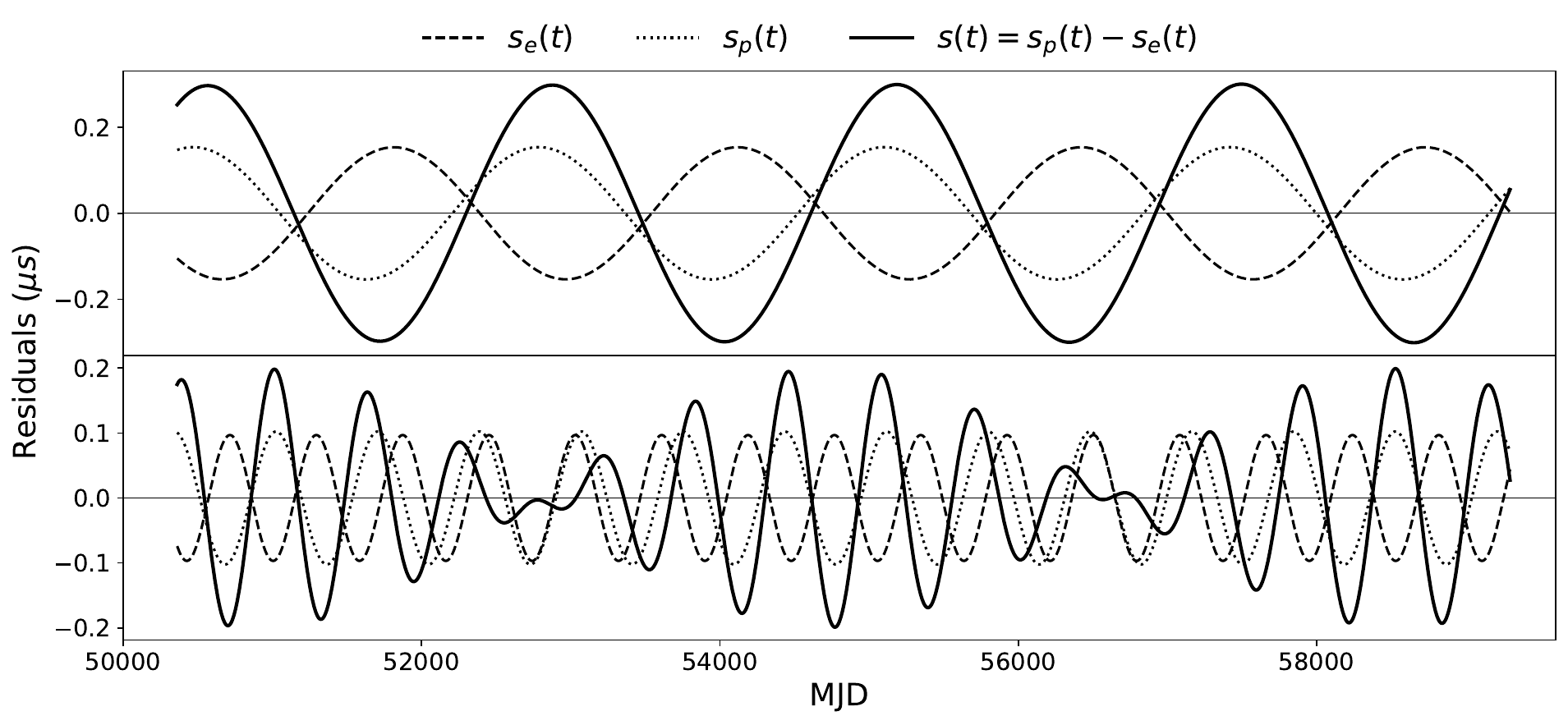}
    \caption{CGWs injected in Pulsar PSR J1022+1001. \textit{Top:} Injected delay induced by the 5 nHz CGW. \textit{Bottom:} Injected delay induced by the 20 nHz CGW. Earth term, pulsar term and total delay are shown.}
    \label{fig:cgw_res}
\end{figure*}

{\it Datasets}. For the purpose of this investigation, seven different datasets were produced, with the exact same noise realization, but different gravitational signals:
\begin{itemize}
    \item {\bf dataset injGWB01, injGWB02, injGWB03}. These three datasets contain a \textit{GWB-only}. They are produced injecting three background selected from the 100 populations mentioned above. The three populations are chosen among those displaying a smooth behaviour in the lowest frequency bins, without particularly loud sources.

    \item {\bf dataset injCGW5, injCGW20}. These two datasets only contain a single loud \textit{CGW-only}. The injection procedure is similar to the GWB datasets, but the population file used contains only one binary. The two injected sources have Earth term GW frequencies of $5$nHz (injCGW5) and $20$nHz (injCGW20). The parameters of the 5nHz source are chosen to approximately reproduce frequency and amplitude of the candidate CGW found in the EPTA DR2new, \cite{CGW_EPTA}, which in fact has a frequency of $\sim$5nHz and an amplitude $h\sim1.1\times 10^{-14}$. The source is then located in the Virgo cluster; this choice determines sky location and distance, and consequently the chirp mass. Inclination, initial phase and polarization angles are randomly extracted as in the GWB case. The source injected in injCGW20 has identical parameters except for the higher frequency (20nHz instead of 5nHz), thus resulting in a higher strain ($h\sim 2.7\times 10^{-14}$). The delays induced in pulsar PSR J1022+1001 by these two CGWs are shown in Fig. \ref{fig:cgw_res}.
    
    \item {\bf dataset injGWB03+CGW5, injGWB03+CGW20}. These two datasets are injected with a GWB {\it and} a CGW. The GWB is in both cases the one resulting from population 03, while the CGWs are either at 5nHz or 20nHz.
\end{itemize}

\begin{table}
      \centering
      \begin{tabular}{|c|c|}
        \hline \textrm{Parameter}&\textrm{Injected value}\\
        \hline \hline
        $\text{cos}\imath$ & -0.44 = cos(2.03 rad)\\
        $\text{cos}\theta$ & 0.21 = cos(12.38°) \\
        $\mathcal{M}$ & $10^{9.06} \sim 1.16\times 10^{9}$M$_{\odot}$\\
        $d_L$ & $10^{1.33} \sim 21.5$ Mpc\\
        $f_{\rm obs}$ & $10^{-8.299} \sim$ 5 nHz \quad / \quad  $10^{-7.699} \sim$ 20 nHZ \\
        $h$ & $10^{-13.97} \sim 1.1\times 10^{-14} \quad / \quad   10^{-13.57}\sim2.7\times 10^{-14}$ \\
        $\Phi_0$ & 3.45 rad\\
        $\phi$ & 3.23 rad = $12^{\rm h}19^{\rm min}50^{\rm s}$\\
        $\psi$ & 0.93 rad \\
        \hline
      \end{tabular}
      \vspace{0.1cm}
      \caption{Parameters of the two injected CGWs. All the parameters are shared except for the frequency, $f_{\rm obs}$, thus resulting in a different strain $h$. Frequency and chirp mass are given in the observer frame, and the luminosity distance is listed for consistency.}
  \label{CGW_parameters}
  \end{table}

\subsection{Signal modelling and analysis methods}\label{analysis}
All the analyses are carried using the \texttt{ENTERPRISE} \citep{enterprise, enterprise_extension} package. The starting point of the analysis is the likelihood function of the timing residuals $\delta \textbf{t}$
\begin{equation}\label{LIke}
    P(\delta \mathbf{t}|\xi, \zeta) = \frac{\text{exp}\left[-\frac{1}{2}(\delta \mathbf{t} - \mathbf{M}\epsilon - \mathbf{s})^T\mathbf{C}^{-1}(\delta \mathbf{t} - \mathbf{M}\epsilon - \mathbf{s})\right]}{\sqrt{(2\pi)^{l}\text{det}\mathbf{C}}}.
\end{equation}
The term $\textbf{M}\epsilon$ is the timing model error expressed as a linear function of the offset from the nominal timing model parameters, $\epsilon$. $\xi$ are the timing model parameters and $\zeta$ are the stochastic components parameters. The likelihood function sampled to search for the gravitational signal is marginalized over the timing model parameters as described in \cite{2022MNRAS.509.5538C}. $\mathbf{C}=\mathbf{C}(\zeta)$ is the covariance matrix that contains all the contributions from stochastic processes, which are:
\begin{itemize}
    \item {\bf WN}, modelled according to Eq.~\eqref{wn}. All the white noise parameters are fixed at the injected values in the GW searches;
    \item {\bf RN}, modelled as a Gaussian process with amplitude of the Fourier components given by Eq.~\eqref{rn_model}  (\cite{2014MNRAS.437.3004L};
    \item {\bf DM}, modelled as a Gaussian process with amplitude of the Fourier components given by Eq.~\eqref{rn_model} and amplitude depending on the observed radio frequency (see Eq.~\ref{DM_freq}). As for RN, the DM is not present in all the pulsars and, if present, the number of Fourier bins used in the Gaussian process depends on the pulsar and it is equal to the number of components used in the injection. For a summary of the noise model characteristics, see Table \ref{noise_model}.
    \item {\bf common red noise (CRN)}, if searched for. A red noise process which is common to all the pulsars can be accounted for in the covariance matrix. If included in the analysis, it is also modelled as a Gaussian process. The amplitude of the components are modelled in two ways: as a power law function of the frequency, and therefore the parameters are the slope $\gamma_{\rm CRN}$ and the amplitude $A_{\rm CRN}$; or as a binned spectrum, and thus the parameters are the amplitudes of the common process in each frequency bin $f_i$. In both cases, the number of components used in the Gaussian process is 30.
    \end{itemize}

Although we inject the GWB as  a superposition of unresolved SMBHBs, we follow the standard PTA analysis practice and model it in our likelihood function as a CRN characterized by an amplitude and a spectral index, as in Eq.~\eqref{rn_model}, which can be directly translated in a characteristic strain of the form given by Eq.~\eqref{eq:hc_phinney}. Note that, depending on the dataset, we use two different flavours of CRN. In most searches, the GWB is modelled as a {\bf common uncorrelated red noise (CURN)}. This means that inter-pulsar correlations induced by the GWB are neglected and the GWB posterior distribution is then reconstructed {\it a posterirori} with the reweighting technique \cite{Hourihane_2023}, when possible. The advantage of this model is that it is much faster since it keeps the correlation matrix in the likelihood function diagonal and it naturally provides the Bayes factor of the reconstructed model versus the  sampled model, besides the posterior distribution of both models. However, the reweighting technique is not always efficient and, as we will see later, using a CURN to model a GWB is not always appropriate; therefore, when needed, we use the {\bf HD-correlated CRN model} which we referred to as GWB model. In this case, GWB induced inter-pulsar spatial correlation are also modelled, according to the HD curve \citep{1983_HD}
\begin{equation}
    \Gamma_\mathrm{ab}=\frac{3}{2} \frac{1-\cos \theta_\mathrm{ab}}{2}
    \ln \left( \frac{1-\cos\theta _\mathrm{ab}}{2} \right)-\frac{1}{4}
    \frac{1-\cos \theta_\mathrm{ab}}{2} +\frac{1}{2}
\end{equation}
where a and b are indexes identifying two pulsars in the array and $\theta_{\rm ab}$ is their angular separation in the sky. Note that searching for an HD-correlated CRN is much more computationally expensive, since the correlation matrix is not diagonal anymore.

The last term in the likelihood function is \textbf{s}, which is the delay induced by an individually resolvable CGW, and is analytically described by Eqs.\eqref{eq:SMBHB_delays}--\eqref{eq:omega} which are implemented in the CGW model of \texttt{ENTERPRISE}. An approximation that can be adopted is to consider in the CGW model the ET only, neglecting the PT. This is computationally very convenient, since the former only depends on the 8 source parameters. Conversely, the latter also depends on the distance of each pulsar in the array. Since those  distances are generally known with large uncertainties (of the order of 20$\%$), they must be considered as additional free parameters in the search. Therefore, the complete model has at least one more parameter per each pulsar with respect to the ET only model. Moreover, as prescribed in \cite{Ellis_2013}, in order to ease the search of the best parameters, one extra parameter is added per each pulsar, which is an initial phase at the pulsar, $\Phi_p$. We will assess the performance of an ET only search on realistic signals including ET and PT in Section \ref{sec:etvspt}.

We mainly adopt Bayesian techniques in our analysis, although we also exploit the frequentist optimal statistics, when useful. The priors used in the search are listed in table \ref{priors} and are all uninformative, except for the pulsar distances' priors that exploit information obtained from independent electromagnetic observations. The prior distributions have a width which is taken as the measurement error listed in \cite{Verbiest_2012}; for the pulsars that are not included in this paper, the relative error is fixed at 20$\%$. The mean of the prior is chosen such that the pulsar distance used for the injection is within 1$\sigma$ of the prior distribution for each pulsar. Parameter estimation is then performed by sampling the posterior distribution with the Markov
Chain Monte Carlo (MCMC) sampler \texttt{PTMCMCSampler}, \cite{justin_ellis_2017_1037579}. For the 25 pulsar PTA we employ, the dimensionality of the posterior distribution is 62 (2 parameters -- slope and amplitude -- for each red noise and dispersion measure process included in the model) plus the GW model dimension (2 for the GWB power-law model, 8 or 58 for an ET only or a full ET+PT CGW model search respectively). Model selection is done by computing the Bayes factor, either with the reweighting technique, when it is efficient, or with the product space method implemented in \texttt{ENTERPRISE\_EXTENSIONS} \citep[for details on the method see][for how it is implemented in PTA]{10.1093/mnras/stv2217, PhysRevD.102.084039}.


\begin{table}
\centering
      \begin{tabular}{|c|c|c|}
      \hline \textrm{Parameter}&\textrm{Prior function}&\textrm{Prior range}\\
      \hline \hline
      ${\rm log}_{10}A_{\rm RN}$ & Uniform & [-11, -18]\\
      $\gamma_{\rm RN}$ & Uniform & [0, 7]\\
      ${\rm log}_{10}A_{\rm DM}$ & Uniform & [-11, -18]\\
      $\gamma_{\rm DM}$ & Uniform & [0, 7]\\
      ${\rm log}_{10}A_{\rm CRN}$ & Uniform & [-11, -18]\\
      $\gamma_{\rm CRN}$ & Uniform & [0, 7]\\
      $\text{cos}\imath$ & Uniform & [-1, 1] \\
      $\text{cos}\theta$ & Uniform & [-1, 1] \\
      $\text{log}_{10}(\mathcal{M}/M_{\odot})$ & Uniform & [6, 10]\\
      $\text{log}_{10}(f/Hz)$ & Uniform & [-9, -7] \\
      $\text{log}_{10}(h)$ & Uniform & [-18, -11] \\
      $\Phi_0$ & Uniform & [0, 2$\pi$] \\
      $\phi$ & Uniform & [0, 2$\pi$] \\
      $\psi$ & Uniform & [0, $\pi$] \\
      $\Phi_p$ & Uniform & [0, $2\pi$] \\
      $L_p$ & $\mathcal{N}(\mu_p, \sigma_p)$ & $(-\infty, +\infty)$ \\
      \hline
      \end{tabular}
      \vspace{0.2cm}
      \caption{Prior distributions' functional form and range for all parameters involved in the analysis.}
      \label{priors}
\end{table}

\section{Results}\label{sec:res}

Using the methods described in the previous section, we carried out several investigations. In Section~\ref{sec:etvspt}, we assess the importance of modelling a CGW including both the Earth and pulsar terms. In Section \ref{GWBvsCGW}, we investigate the degeneracy between the analysis results obtained with a GWB-only dataset and with a CGW-only dataset. In Section \ref{CGWasGWB}, it is shown how CGW-only datasets can produce strong evidence in favour of an HD-correlated GWB. Viceversa, in Section \ref{GWBasCGW}, it is shown how a GWB-only dataset can be misinterpreted as a continuous wave. This degeneracy is solved in Section \ref{combined}, by means of a combined analysis that searches simultaneously for a GWB and a CGW on both GWB-only and CGW-only datasets . Finally, in Section 
\ref{combined_searchanddata}, we evaluate the performance of a combined analysis of a GWB and a CGW on datasets containing both signals.

\subsection{On the importance of the pulsar term in CGW modelling}\label{sec:etvspt}


\begin{figure*}
\begin{minipage}{.5\linewidth}
\centering
{\label{a}\includegraphics[scale=.25]{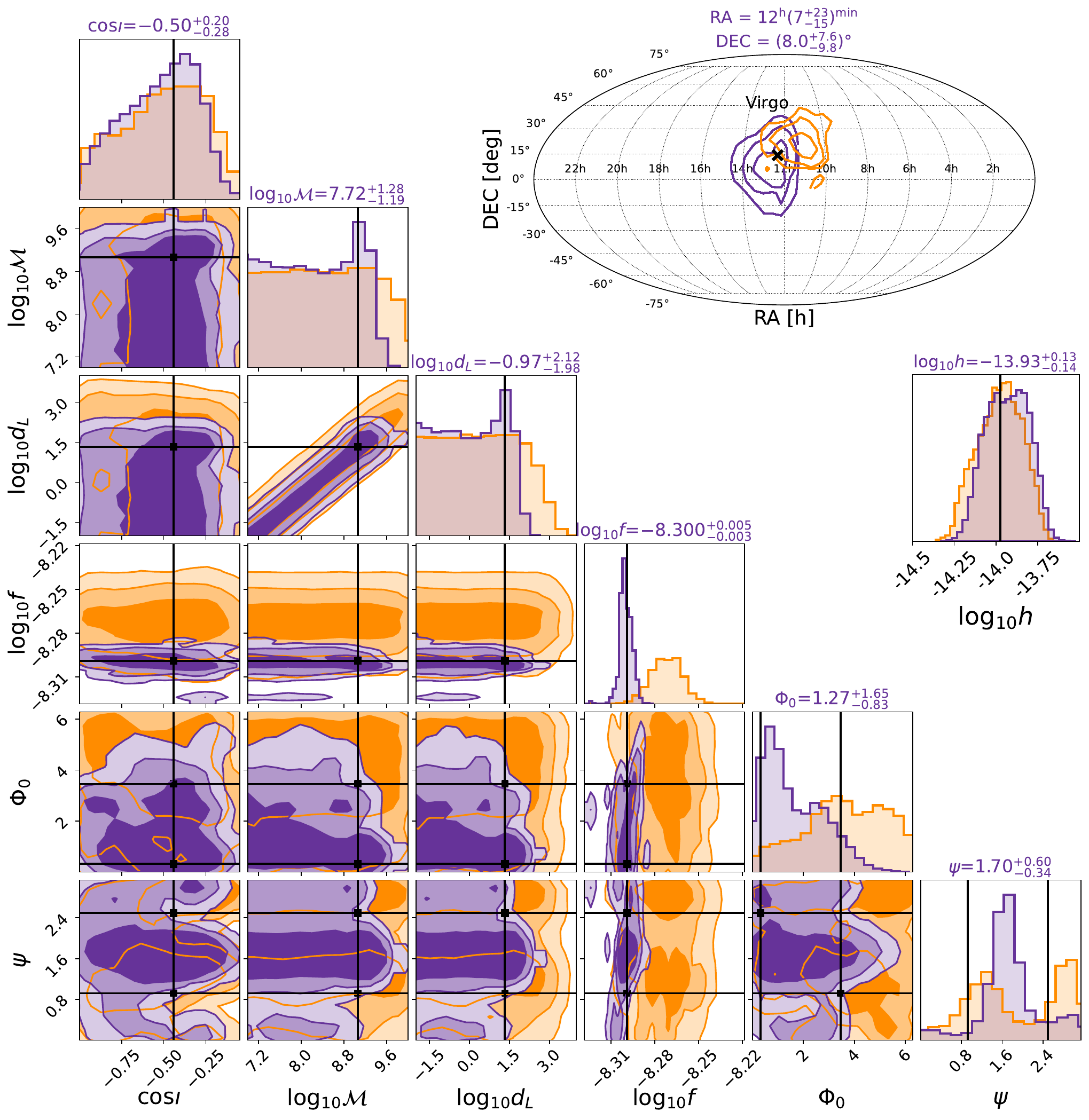}}
\end{minipage}
\begin{minipage}{.5\linewidth}
\centering
{\label{main:b}\includegraphics[scale=.25]{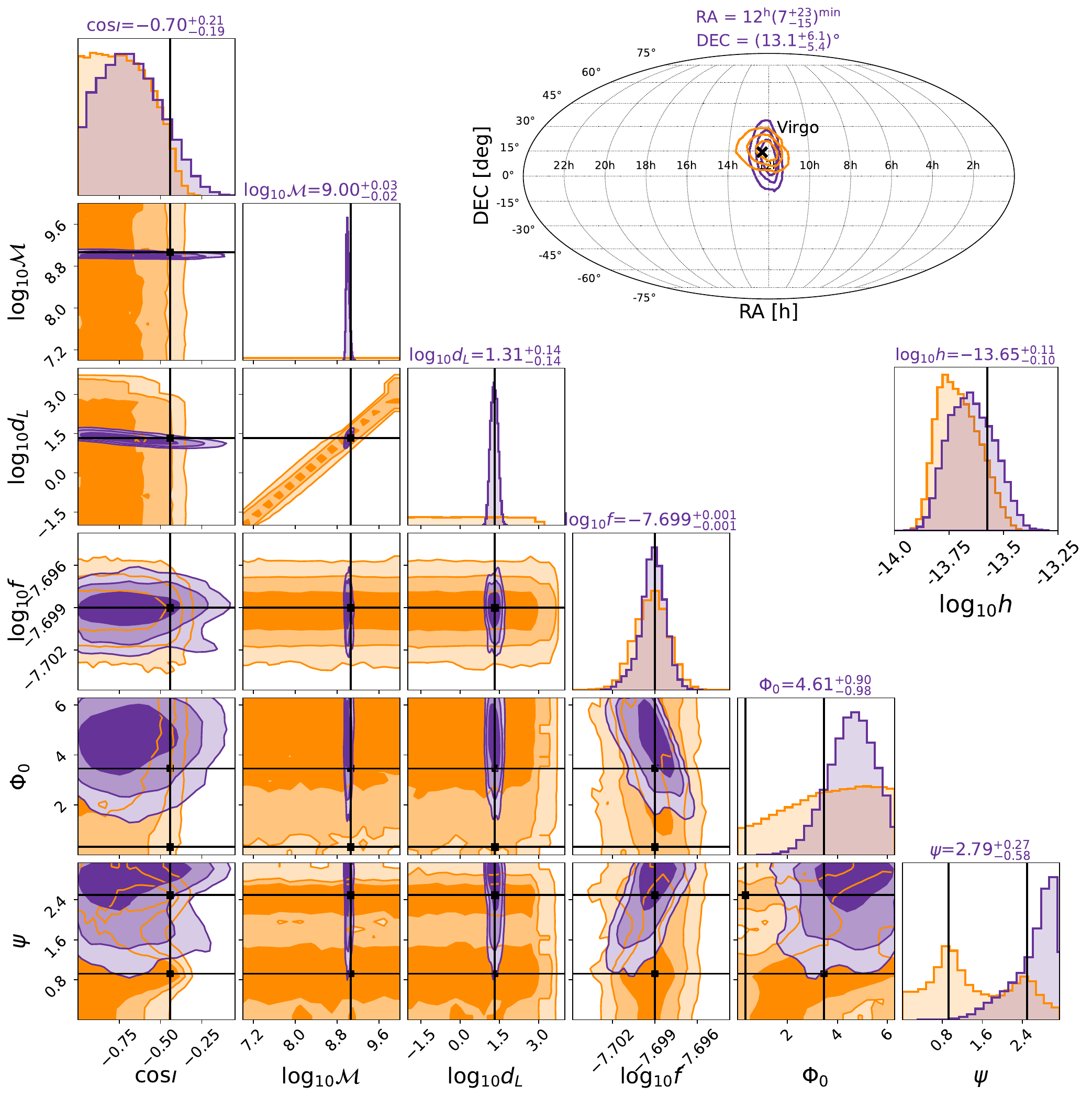}}
\end{minipage}
\caption{Marginalised posterior distributions of the 8 source parameters (6 in the corner plot plus the 2 sky coordinates in the skymap) and of the CGW amplitude (obtained as combination of distance, frequency and chirp mass posterior distributions, {\it stand-alone panel}) for the CGW injected in the dastaset  injCGW5 ({\it left panel}) and injCGW20 ({\it right panel}). The signal is searched with a CGW-only model with ET-only (orange distributions) and with the complete ET+PT template (purple distributions). Injected values are shown as black solid lines.}
\label{fig:etvspt}
\end{figure*}

\begin{figure*}[ht]
   \centering
   \includegraphics[width=0.8\textwidth]{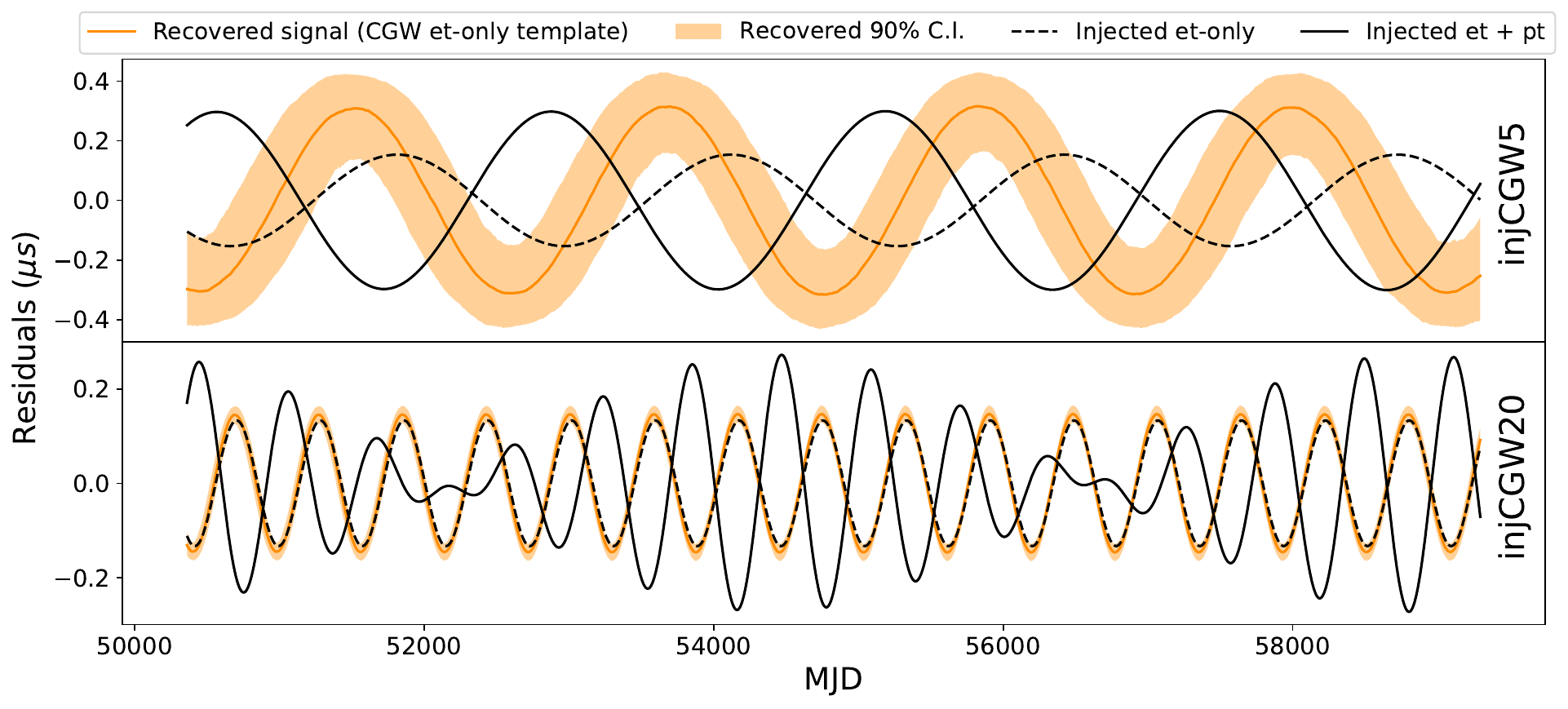}
   \caption{Signal recovered with the ET-only template applied to the 5nHz CGW (dataset injCGW5, {\it top panel}) and to the 20nHz CGW (data injCGW20, {\it bottom panel}). Median and 90$\%$ credible interval of the signal recovery are shown for pulsar PSR J1022+1001 ({\it top}) and PSR J1713+4707 ({\it bottom}). Solid and dashed lines are respectively the injected total and the ET-only signals.}
   \label{fig:cw520_etres}%
\end{figure*}
\begin{figure*}[ht]
   \centering
   \includegraphics[width=0.8\textwidth]{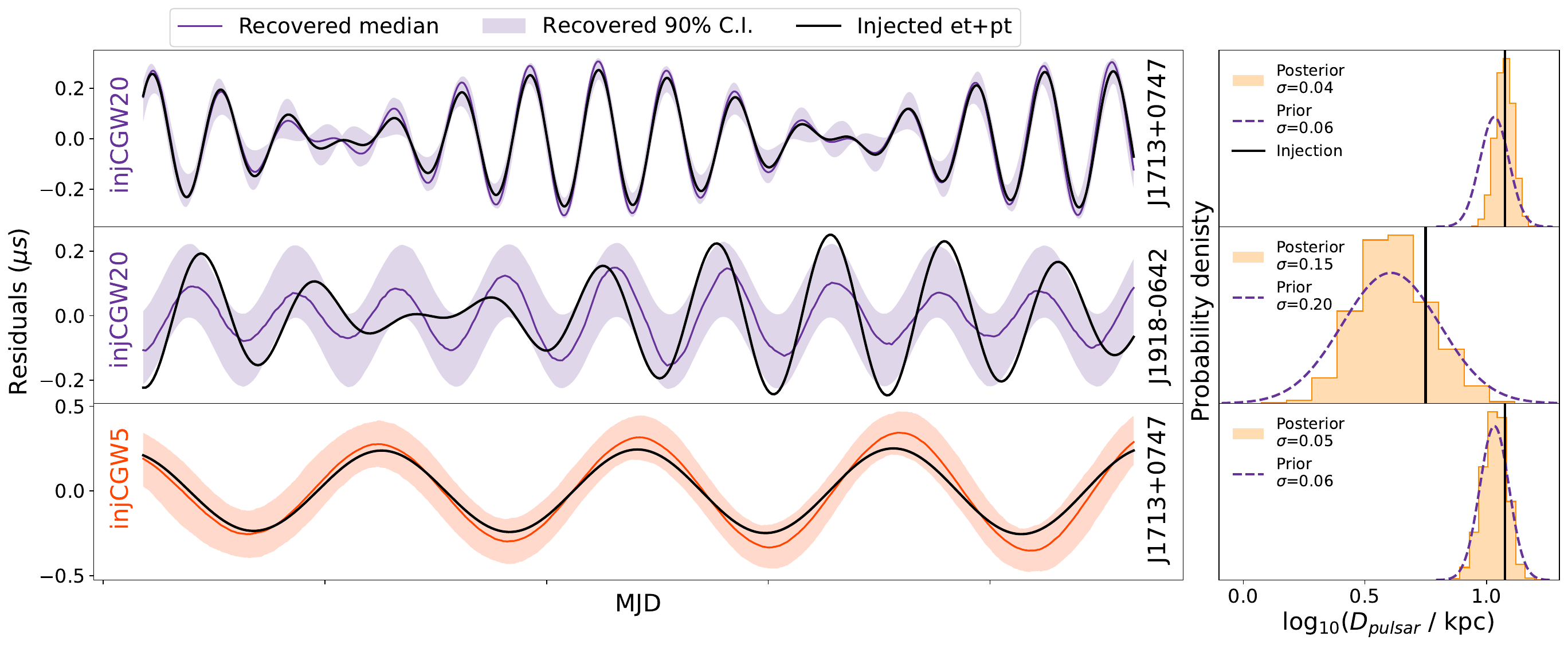}
   \caption{The first column shows the injected CGW residuals (black solid line) and the reconstructed residuals with median and 90$\%$ C.I. (solid line and shaded area). The second column shows the pulsar distance  prior distribution (dashed line) and recovered posterior distribution (histogram) and the injected value of the distance (solid black line). The first two rows correspond to the results obtained by analysing injCGW20 dataset and are shown for pulsar PSR J1713+0747 ({\it first row}) and pulsar PSR J1918-0642 ({\it second row}); the last row shows the results obtained with injCGW5 dataset displayed for pulsar PSR J1713+0747.}
   \label{fig:pulsar_dist}%
\end{figure*}
\begin{figure}[ht]
   \centering
   \includegraphics[width=0.4\textwidth]{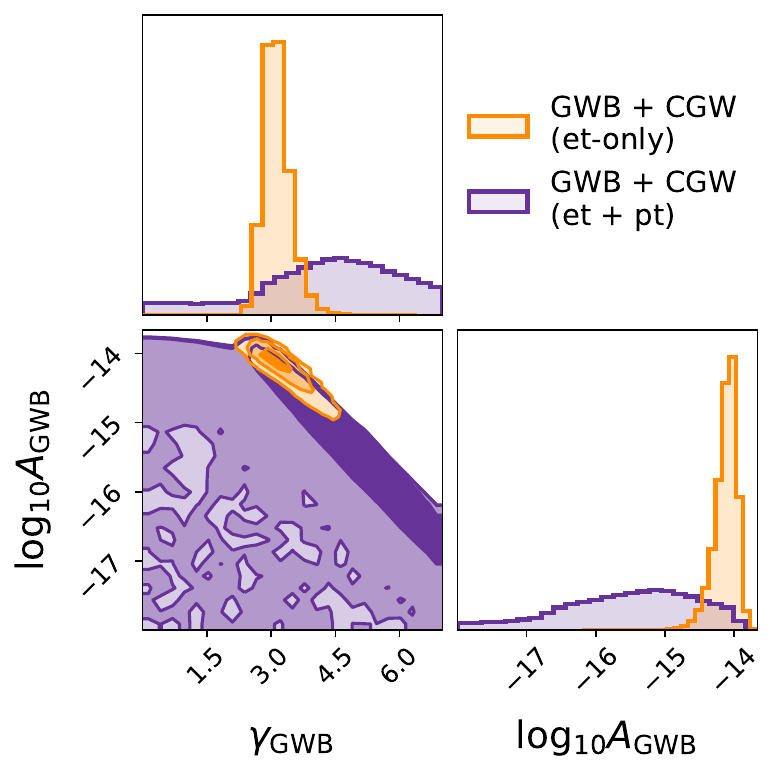}
   \caption{Posterior distribution of the GWB parameters obtained from dataset injCGW5 with the GWB+CGW ET-only (purple) and GWB+CGW ET+PT (orange) template.}
   \label{fig:GWB_etvspt}%
\end{figure}

As described in section \ref{analysis}, the signal imprinted by a CGW in the ToAs is the sum of two terms: the Earth term and the pulsar term. The ET is common to all pulsars and only depends on the 8 parameters describing the source. Conversely, the PT depends on the distance of each pulsar and the phase of the GW signal at the pulsar's location. 
This means that modelling of the full CGW ET+PT in a search performed on an array of 25 pulsars (like the one we are considering here) requires 50 extra free parameters, that must be added to the 8 source parameters and to the 62 parameters of the individual pulsars noise components. It is therefore interesting to compare searches of CGW modelled as ET only and ET+PT on realistic datasets, to see whether the computational efficiency of the ET only search bears any significant drawback.

The two templates have been tested on both CGW-only datasets (injCGW5 and injCGW20), and the main results are shown in Fig.~\ref{fig:etvspt}. From the figure it is clear that the performance of the ET only template depends on the CGW frequency. In fact, when applied to the injCGW20 dataset (right panel of Fig.~\ref{fig:etvspt}), the ET only model performs well on all parameters that it is sensitive to, i.e. all source parameters except the chirp mass and the luminosity distance. Conversely, the same model applied to the injCGW5 dataset results in some non-negligible biases in the recovery. As it can be seen from the left panel of Fig.~\ref{fig:etvspt}, the injected values for the frequency and sky location are not compatible with the $90\%$ C.I. of the recovered posterior distributions. 

The different performance of the ET only template in the two cases is due to the different degree of separability of the ET and PT terms for the two injected sources. For the injected system chirp mass, $\mathcal{M}=1.16\cdot 10^9M_{\odot}$, the average frequency difference between ET and PT is $\overline{\Delta f} \sim 
3.8\times10^{-6}$nHz$\,$yr$^{-1}\overline{\Delta t}\sim 1.0\times 10^{-2}{\rm nHz} \sim 0.02\Delta f_{bin}$ for the 5nHz CGW and $\overline{\Delta f} \sim 1.4\times 10^{-3}$nHz$\, $yr$^{-1}\overline{\Delta t}\sim 4.0{\rm nHz} \sim 3.2\Delta f_{bin}$ for the 20nHz CGW. Therefore, while in the 20nHz CGW, ET and PT falls at very distinct frequencies, they blend together in the same frequency bin for the 5nHz CGW. In the former case, the ET only template only picks the common power produced by the ET at its frequency, disregarding the PT term; while in the latter, the ET only template is trying to fit a superposition of ET and PT essentially overlapping at the same frequency. This concept is demonstrated in Fig. \ref{fig:cw520_etres}, where the reconstructed waveforms are compared to the injected ones. In the 20nHz case (injCGW20 dataset), the ET only template almost perfectly fits the ET of the full injected waveform, while in the 5nHz case (injCGW5), the ET only template absorbs power from both ET and PT and the reconstructed waveforms sits, for many pulsars, somewhat in between the ET and the full injected signal and is also way more uncertain as shown by the much larger credible region.



Conversely, the full ET+PT template is able to correctly reconstruct the signal and, because it is sensitive to the frequency evolution, it also allows to constrain both the source chirp mass and the luminosity distance. The performance is very different for the 5 nHz CGW and the 20 nHz CGW: for the former the chirp mass recovery is $\mathcal{M}=0.52^{+9.98}_{-0.48}\times 10^8M_{\odot}$, while for the latter it is $\mathcal{M}=1.01^{+0.07}_{-0.06}\times10^9M_{\odot}$, where we give the median and the 68$\%$ credible interval. The superior recovery of the chirp mass in the 20 nHz case is due to the much larger frequency evolution of the source between ET and PT, allowing to place a much tighter constrain on $\dot{f}$ and thus on ${\cal M}$.
Another consequence of the large frequency evolution of the 20 nHz CGW is that the model is sensitive to the pulsar distances. This is shown in Fig.\ref{fig:pulsar_dist} where the reconstructed waveform (left) and the posterior distribution of the pulsar distance (right) are shown for PSR J1713+0747 for the injCGW20 (top row) and injCGW5 (bottom row) datasets. In the first case, the posterior distribution on the distance is narrower than the prior and centred at the injected value: this is allowed by a correct modelling of the frequency evolution, which is reflected in the excellent reconstruction of the waveform. Conversely, for the 5 nHz case, the posterior distribution of the pulsar distance is almost indistinguishable from the prior, but signal reconstruction still matches the injection, as the frequency evolution is too small for the search to be sensitive to it. We note that the GW signal reconstruction and distance determination is not always as good for the 20 nHz CGW. This is shown in the middle row of Fig.\ref{fig:pulsar_dist} for pulsar PSR J1918-0642. In this case the posterior distribution resembles the prior, and while the reconstructed signal does not strongly disagree with the injection, the median fails to capture the signal shape accurately. In fact, only about 13 pulsars, whose evolution is accurately reconstructed, significantly contribute to the chirp mass recovery.


In addition to better and unbiased recovery of the CGW parameters, there is another rationale for preferring the complete CGW template over the ET-only template: in a joint CGW + GWB search, the contribution to the CGW signal due to the pulsar term, if not taken into account by the CGW model, is absorbed by the GWB model, since it is a common red noise process. It will therefore add extra spurious power to the GWB, biasing the recovery of its parameters. An example of this behaviour is shown in Fig.~\ref{fig:GWB_etvspt}, where the GWB+CGW search is applied to the dataset injCGW5: when the CGW is searched with the full template, the posterior distribution of the GWB parameters is unconstrained, when the search is done with the ET-only template, the posterior distribution of the GWB parameters remains well constrained because it absorbs all the power coming from the unmodelled pulsar term.

We therefore conclude that while the ET-only template can be used to search for possible CGW candidates and provides a reasonable estimate of the source parameters \citep[see also][]{2024PhRvL.132f1401C}, the full ET+PT model must be preferred, especially in a joint search with a stochastic GWB. Despite the much lower computational efficiency, we will therefore always model our CGWs as ET+PT in the following.

\subsection{Degeneracy between GWB and CGW}\label{GWBvsCGW}
In this section, searches for GWB, CGW and combined searches of the two signals together are applied to the 3 GWB realizations (injGWB01, injGWB02, injGWB03) and to the 2 CGW realizations (injCGW5, injCGW20). The aim is to study the degeneracy between the two signals and how it can be solved.

\subsubsection{CGW misinterpreted as a GWB}\label{CGWasGWB}

We start by performing a GWB analysis on the three injGWB and the two injCGW datasets. To this end we make use of both Bayesian and frequentist techniques following this procedure:
\begin{itemize}
    \item a power law-shaped and HD-correlated stochastic process is assumed for the GWB, then the slope $\gamma_{\rm GWB}$ and the amplitude log$_{10}A_{\rm GWB}$ of the powerlaw process are sampled, together with the noise parameters to obtain the posterior distribution;
    \item the Bayes factor $\mathcal{B}^{\rm GWB}_{\rm CURN}$ is evaluated to compare the evidence for an HD-correlated and an uncorrelated red noise process. The bayes factor is computed either with the reweigthing method or with the product space when reweighting is not efficient;
    \item the averaged angular correlation between the pulsars is estimated with the optimal statistics method averaging over 10 angular bins, each one containing 30 pulsar pairs. A powerlaw-shaped background with variable $\gamma$ is assumed;
    \item the SNR of the HD cross-correlation marginalized over the noise parameters is obtained with the optimal statistics method, assuming a powerlaw-shaped background with variable $\gamma$.
\end{itemize}

\def\arraystretch{1.4} 
\begin{table*}
\centering
      \begin{tabular}{|c|c|c|c|c|}
      \hline \textrm{Dataset}&$\mathcal{B}^{\rm GWB}_{\rm CURN}$& SNR ($\gamma$) & log$_{10}A_{\rm GWB}$ & $\gamma_{\rm GWB}$\\
      \hline \hline
      injGWB01 & $9600 \pm 500$ & $6.9_{-1.4}^{+1.3}\quad (13/3)$ & $-14.73_{-0.36}^{+0.30}$ & $4.55_{-0.68}^{+0.66}$\\ 
      injGWB02& $106.7 \pm 0.59$ & $3.6_{-0.6}^{+0.7} \quad(13/3)$ & $-14.20_{-0.19}^{+0.13}$ & $3.50_{-0.27}^{+0.35}$\\
      injGWB03 & $0.156 \pm 0.003$ & $0.4_{-0.5}^{+0.5}\quad (13/3)$& $-14.51_{-0.15}^{+0.13}$ & $4.15_{-0.28}^{+0.29}$\\
      injCGW5 & 10 & $3.0_{-0.5}^{+0.5}\quad (3)$ & $-14.01_{-0.10}^{+0.09}$ & $3.09_{-0.21}^{+0.23}$\\
      injCGW20 & 60 & $3.2_{-0.6}^{+0.6}\quad (2)$ & $-13.53_{-0.05}^{+0.05}$ & $1.96_{-0.16}^{+0.15}$\\
      \hline
      \end{tabular}
      \vspace{0.2cm}
      \caption{GWB recovery performed on datasets including only a GWB or a CGW. Columns are: dataset name, HD-correlated versus uncorrelated common red noise Bayes factor, HD S/N, median and 68$\%$ credible region of the GWB amplitude and slope.}
      \label{HDvsCURN}
\end{table*}
\begin{figure}
   \centering
   \includegraphics[width=0.5\textwidth]{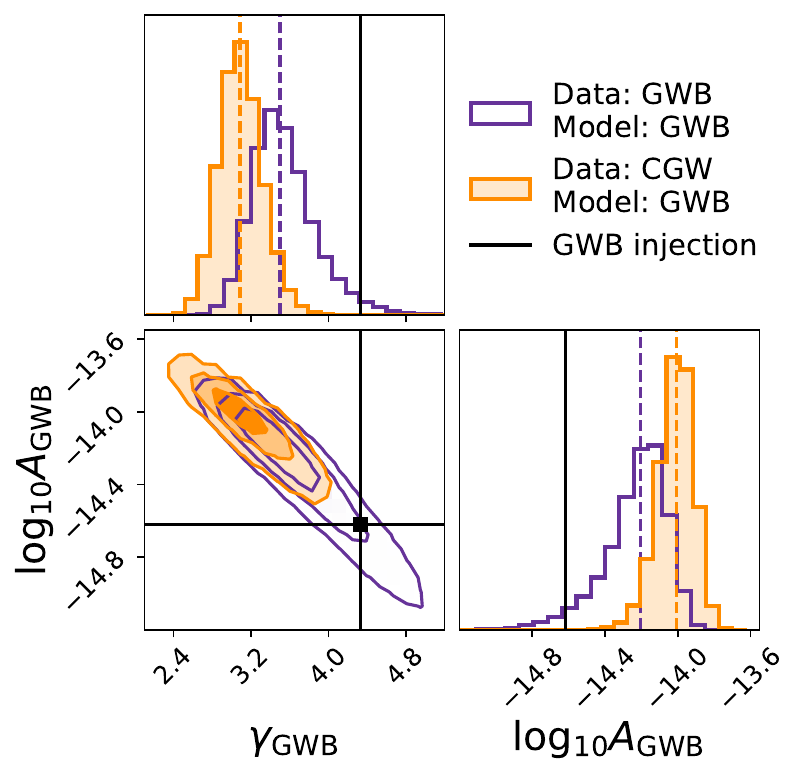}
   \caption{Posterior distribution of the powerlaw GWB parameters. Purple empty posterior is obtained from dataset injGWB02, orange full posterior is obtained from dataset injCGW05. The solid line is the 'nominal' GWB injection ($\gamma\sim$13/3, log$_{10}$A$_{\rm GWB}\sim$ = -14.66).}
   \label{fig:GWB}%
\end{figure}

\begin{figure}
    \centering
    \includegraphics[width=0.5\textwidth]{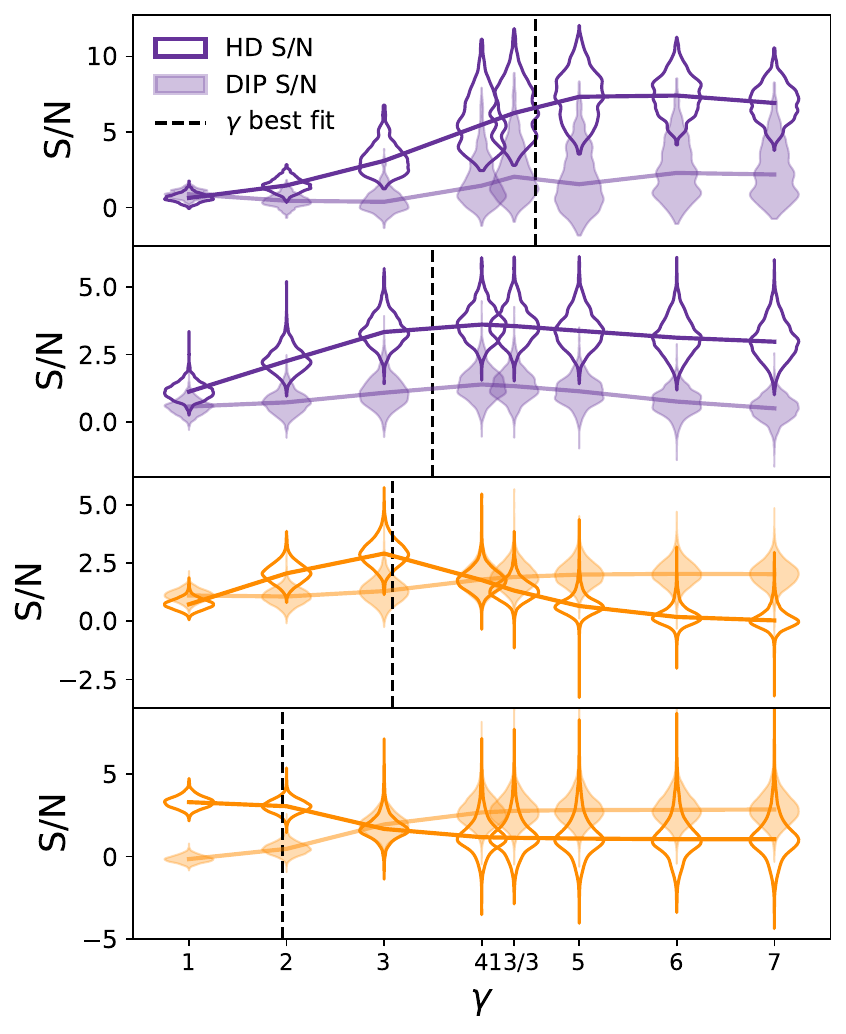}
    \caption{Optimal statistics S/N for an HD-correlated (empty violins) and dipole-correlated (filled violins) process as a function of the assumed $\gamma$ power law index. The two upper panel show in purple the results for datasets injGWB01 ({\it top}) and injGWB02 ({\it bottom}), while the two lower panels show in orange the results for datasets injCGW5 ({\it top}) and injCGW20 ({\it bottom}).}
    \label{fig:HDDIP_SNR}
\end{figure}

The results of the powerlaw spectrum parameter estimation are listed in the last two columns of Table \ref{HDvsCURN} for all five datasets and two examples of the recovered GWB parameters' posterior distributions are shown in Fig.~\ref{fig:GWB} for datasets injGWB02 and injCGW5. In general, the GWB search returns well constrained posteriors regardless of the actual content of the dataset, which in practice means that a CGW can easily be detected as (and mistaken for) a GWB. However, the signal recovered from the two CGW-only datasets (injCGW5, injCGW20) is higher and flatter than the one obtained from the background datasets. This can be clearly seen both from the $A$ and $\gamma$ values of Table \ref{HDvsCURN} and from the 2D contours in Fig.~\ref{fig:GWB}. This is expected since the extra power due to the two single sources is at frequencies which are higher than the loudest GWB frequency bins that dominate the powerlaw spectrum recovery in the GWB-only datasets.



The Bayes factors reported in Table \ref{HDvsCURN} show that an HD-correlated common process (i.e. a GWB) is generally strongly favoured regardless of the content of the data. In fact, dataset injCGW5 produces a Bayes factor $\mathcal{B}^{\rm GWB}_{\rm CURN}$ of $\sim 10$ and HD S/N of 3.0$^{+0.5}_{-0.5}$ (fixing $\gamma$ to 3), while injCGW20 
gives $\mathcal{B}^{\rm GWB}_{\rm CURN}$ of $\sim 60$ and S/N = $3.2^{+0.6}_{-0.6}$ (fixing $\gamma$ to 2). Interestingly, injGWB03 does not favour an HD correlated process, displaying $\mathcal{B}^{\rm GWB}_{\rm CURN}$ of $\sim 0.15$. This is consistent with \cite{2024A&A...683A.201V}, who found that, due to the stochastic nature of the signal, individual realizations of the same SMBHB population can result in Bayes factor that differ by several orders of magnitude.

\begin{figure}
   \centering
   \includegraphics[width=0.5\textwidth]{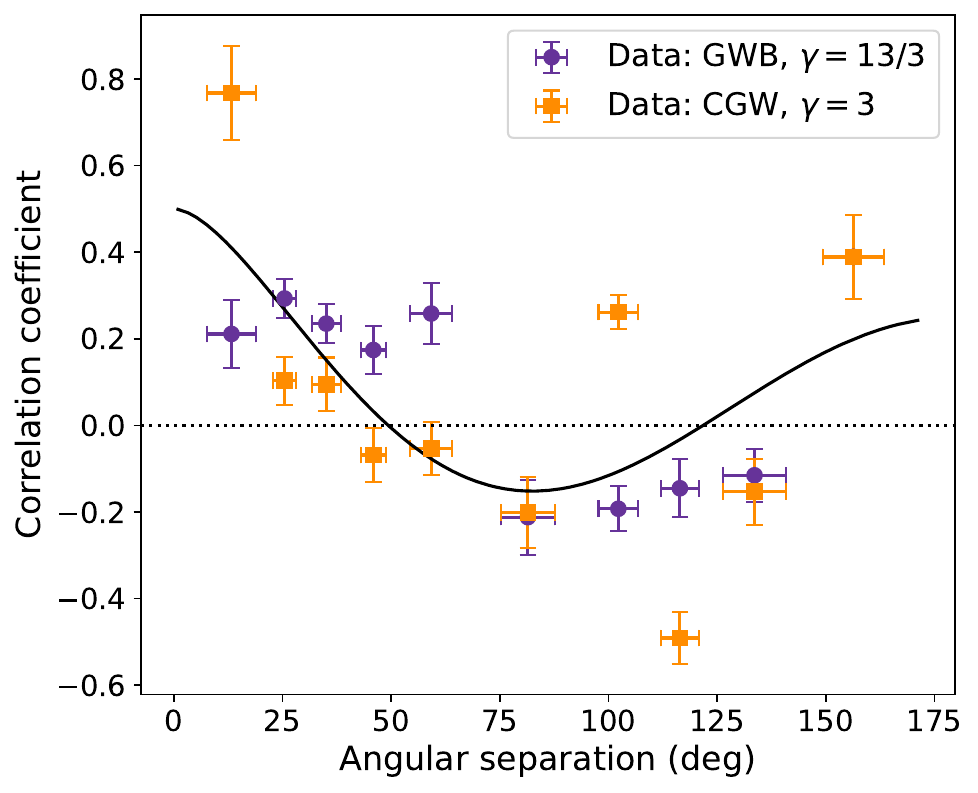}
   \caption{Overlap reduction function obtained with the optimal statistic. Purple circular and orange squared points indicate the results for the injGWB02 and injCGW5 datasets respectively. The correlation coefficients for each pair of pulsars are grouped in 10 bins of 30 pulsar pairs each, the points represent the weighted average in each bin. The solid line shows the expectation value of the HD correlation. Note that the purple rightmost point lie below the orange one.}
   \label{fig:OS}
\end{figure}

Although the Bayes factors obtained on the  injCGW5 and injCGW20 are compatible with a GWB detection, the resulting HD S/N is significant only in a small range of values of $\gamma$ that are consistent with the posterior distributions obtained by the Bayesian search, as shown in Fig.~\ref{fig:HDDIP_SNR}. By fixing $\gamma$ to the nominal value of 13/3, the HD S/N drops to 1.3$^{0.4}_{-0.4}$ and to 1.1$^{1.0}_{-0.9}$ for the injCGW5 and injCGW20 datasets respectively. Conversely, the 
injGWB01 and injGWB02 datasets (also shown in Fig.~\ref{fig:HDDIP_SNR}) result in distributions of the HD S/N that are consistently high for all values of $\gamma\geq 4$.
Moreover, at higher values of $\gamma$, the CGW-only dataset has an increasing evidence for a dipolar-correlated signal, whose S/N surpasses that of HD. This behaviour is not observed in the GWB-only datasets, for which the dipole S/N always stays lower than the HD S/N. Finally, two examples of HD reconstruction are shown in Fig.~\ref{fig:OS} for datasets injGWB02 and injCGW5. While the quadrupolar pattern can be seen in both datasets, the scatter around the expected relation appears to be much larger when a CGW only is present in the data, consistent with \cite{Cornish_2013} and \cite{Allen_2023}. Moreover, the HD correlation for the injCGW5 dataset is reconstructed assuming $\gamma=3$, while it tends to disappear for higher values of $\gamma$.

\begin{figure*}
   \centering
   \includegraphics[width=0.7\textwidth]{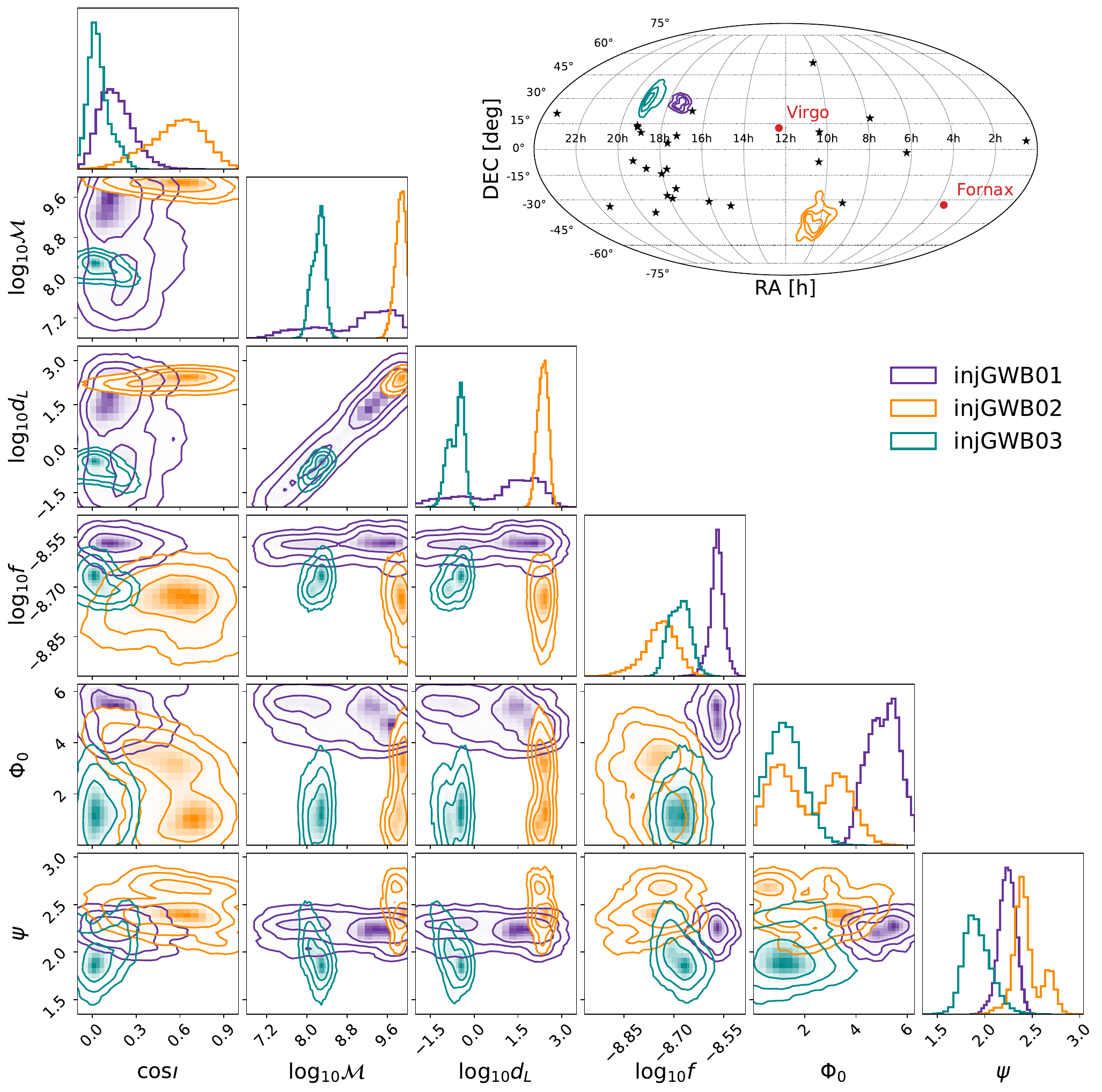}
   \caption{Posterior distributions of the CGW parameters -- inclination angle, chirp mass, luminosity distance, frequency, initial phase and polarization angle -- recovered with a CGW-only model (considering both Earth and pulsar term) applied to the three GWB only datasets. The inset sky map shows the reconstruction of the (fake) CGW sky location. The black stars in the sky map represent the positions of the 25 EPTA pulsars, while the two red dots are the Virgo and the Fornax clusters, there as a reference.}
   \label{fig:1081_CGW}%
\end{figure*}
\begin{figure*}
   \centering
    \includegraphics[width=1\textwidth]{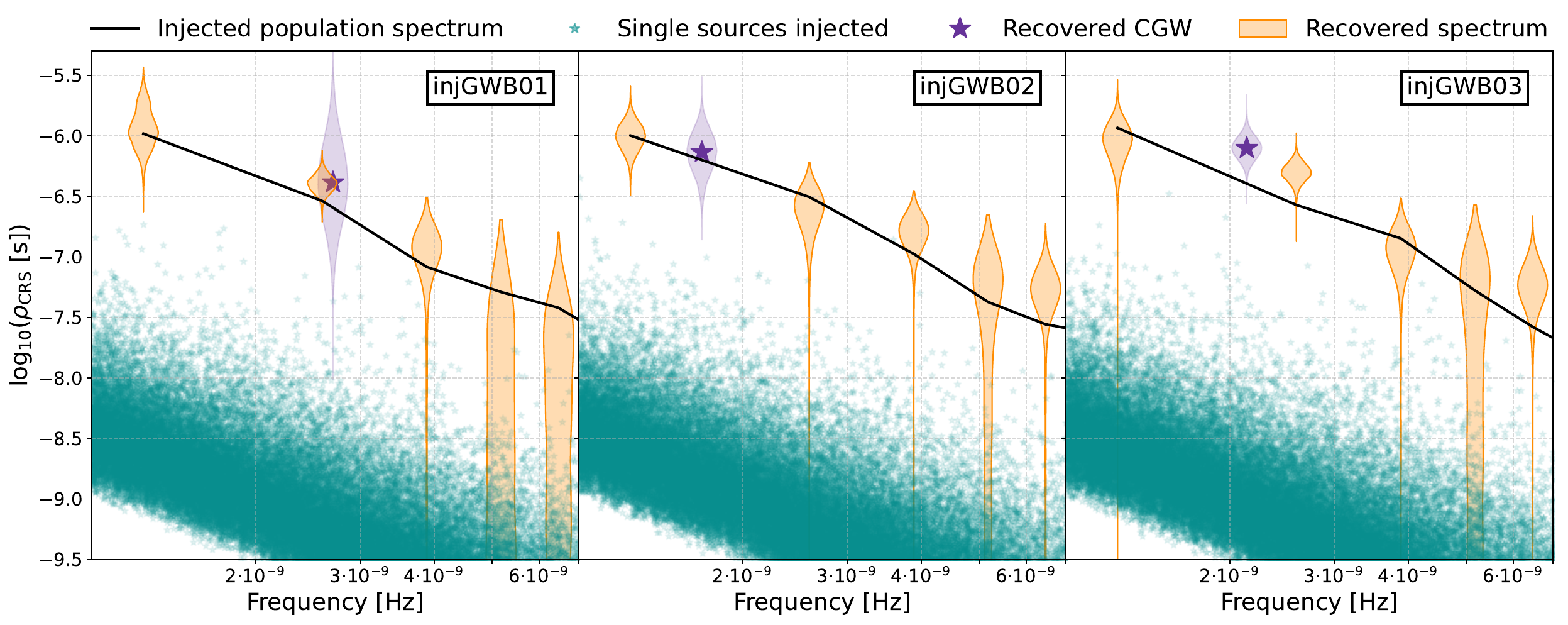}
   \caption{Injected and recovered signals in the three injGWB datasets. In each panel. the green stars represent the contributions of each binary of the injected population to the GWB, the black line is the total power spectral density per frequency bin. The orange violin plot is the HD-correlated free spectrum recovery and the purple violin is the posterior distribution of the recovered CGW amplitude at the location of the recovered CGW frequency.}
   \label{fig:fake_cgw1}%
\end{figure*}

\subsubsection{GWB misinterpreted as a CGW}\label{GWBasCGW}

We now search for a CGW in the same 5 datasets analysed in the previous section. In the search, we estimate the posterior distributions of all the noise parameters -- except for the WN, which is fixed -- and the CGW parameters -- 8 source parameters plus 50 additional parameters for the pulsars' distances and phases. We then use hypermodeling to evaluate the Bayes factor between a model containing a CGW versus a model featuring intrinsic pulsar noise only. 

We can see from Table \ref{cgw_vs_psrn} that also in this case, regardless on whether the dataset contains a CGW or a GWB, a CGW is recovered with high confidence, with a Bayes factor $\mathcal{B}^{\rm CGW}_{\rm PSRN}>1000$ in all cases. Table \ref{cgw_vs_psrn} also shows that the CGW frequency and amplitude are fairly well constrained. The performance of the CGW search on the injCGW datasets has already been assessed in Section \ref{sec:etvspt}. As shown in Fig.~\ref{fig:etvspt} when a full ET+PT template is used for the CGW, the model can correctly recover the injected signals and estimate their parameters, although with differences depending on the frequency of the source. Here it is particulary interesting to check the CGW recovery on the three injGWB datasets, which is shown in Fig. \ref{fig:1081_CGW}: all source parameters are fairly well constrained, including the sky location, which is different in each dataset. These parameters, however, do not correspond to any specific source present in the datasets, as can be seen in the three panels of Fig. \ref{fig:fake_cgw1}. The recovered CGW frequency always corresponds to the frequency bin that is the most constrained by the HD-correlated free spectrum search; moreover, the amplitude recovered is always compatible with the GWB amplitude in that frequency bin. Thus, it can be concluded that the CGW parameters are constrained because the CGW model absorbs part of the HD-correlated common red noise produced by the incoherent superposition of all the SMBHBs composing the GWB. 

Finally, we note that the CGW amplitudes recovered from the three injGWB datasets are respectively $1.2^{+1.5}_{-0.7}\times 10^{-14}$, $1.4^{+0.6}_{-0.5}\times 10^{-14}$  and $1.8^{+0.5}_{-0.4}\times 10^{-14}$ (median and 68$\%$ credible interval). These values are all well compatible with the amplitude of the DR2new CGW candidate, which is $1.0^{+0.5}_{-2.6}\times 10^{-14}$ (median and 90$\%$ credible interval) \citep{CGW_EPTA}.

\begin{table}
\centering
      \begin{tabular}{|c|c|c|c|}
      \hline \textrm{Dataset}&$\mathcal{B}^{\rm CGW}_{\rm PSRN}$&\shortstack{recovered \\ CGW frequency}&\shortstack{recovered \\ CGW amplitude}\\
      \hline \hline
      injGWB01 & > 1000 & $2.70^{+0.05}_{-0.16}$ nHz & $-13.91^{+0.35}_{-0.35}$\\
      injGWB02 & > 1000 & $1.7^{+0.2}_{-0.2}$ nHz & $-13.87^{+0.18}_{-0.22}$\\
      injGWB03 & > 1000 & $2.1^{+0.1}_{-0.1}$ nHz & $-13.73^{+0.11}_{-0.13}$\\
      injCGW5 & > 1000 & $5.01^{+0.06}_{-0.03}$ nHz & $-13.93_{-0.14}^{+0.13}$\\
      injCGW20 &  > 1000 & $19.99^{+0.05}_{-0.05}$ nHz & $-13.65^{+0.11}_{-0.10}$\\
      \hline
      \end{tabular}
      \vspace{0.2cm}
      \caption{CGW recovered from datasets containing only a GWB or a CGW. Columns are dataset name, Bayes factor of PSRN + CGW versus PSRN only, median and 68$\%$ credible region of the recovered CGW frequency and amplitude.}
      \label{cgw_vs_psrn}
\end{table}


\begin{table*}
\centering
      \begin{tabular}{|c|c|c|c|c|c|c|c|}
      \hline \textrm{Dataset}&$\mathcal{B}^{\rm CURN+CGW}_{\rm CURN}$&$f_{\rm cgw}$ [nHz]&log$_{10}h_{\rm cgw}$&$\mathcal{B}^{\rm GWB+CGW}_{\rm GWB}$&log$_{10}A_{\rm GWB}$ & $\gamma_{\rm GWB}$&log$_{10}h_{95\%}$\\
      \hline \hline
      injGWB01 & 3.6 & $2.48_{-1.21}^{+14.01}$& $-14.12_{-0.84}^{+0.63}$& 0.75 & $-14.71_{-0.37}^{+0.30}$ & $4.49_{-0.70}^{+0.65}$ & -14.19\\
      injGWB02&  1.0 & $3.42_{-1.22}^{+23.58}$& $-14.29_{-0.93}^{+0.19}$& 0.7 & $-14.16^{+0.11}_{-0.19}$ & $3.41^{+0.35}_{-0.24}$ & -13.92\\
      injGWB03 & 1.3 & $4.71_{-2.57}^{+31.31}$& $-14.40_{-1.11}^{+0.62}$& 0.6 & $-14.46_{-0.16}^{+0.15}$ & $4.07_{-0.30}^{+0.32}$ & -13.71\\
      injCGW5 & > 1000 & $5.07_{-0.06}^{+0.04}$& $-13.90_{-0.16}^{+0.14}$& > 1000 & $-15.57_{-1.17}^{+0.99}$ & $4.33_{-2.19}^{+1.59}$ & -\\
      injCGW20 & > 1000 & $20.04_{-0.06}^{+0.08}$& $-13.66_{-0.10}^{+0.11}$& > 1000 & $-14.09_{-1.37}^{+0.19}$ & $2.84_{-0.45}^{+1.81}$ & -\\
      \hline
      \end{tabular}
      \vspace{0.2cm}
      \caption{Summary of the CURN+CGW and GWB+CGW searches results on all five datasets containing either a GWB or a CGW. Columns are: 1) dataset to which the search is applied, 2) Bayes factor of CURN+CGW vs CURN models, 3) median and 68$\%$ credible interval of the CGW frequency and 4) CGW amplitude recovered with the CURN+CGW model, 5) Bayes factor of GWB+CGW vs GWB models, 6) median and 68$\%$ of the GWB amplitude and 7) GWB slope recovered with GWB+CGW model, 8) $95\%$ upper limit on the CGW amplitude recovered with the GWB+CGW model. The last column is not reported for the injCGW datasets, for which the CGW was detected and the posterior on its amplitude well constrained.}
      \label{gwbcw_vs_gwb}
\end{table*}

\subsubsection{Breaking the degeneracy: combined search for a GWB and a CGW}\label{combined}

\begin{figure*}
\begin{minipage}{.5\linewidth}
\centering
{\label{a}\includegraphics[scale=.37]{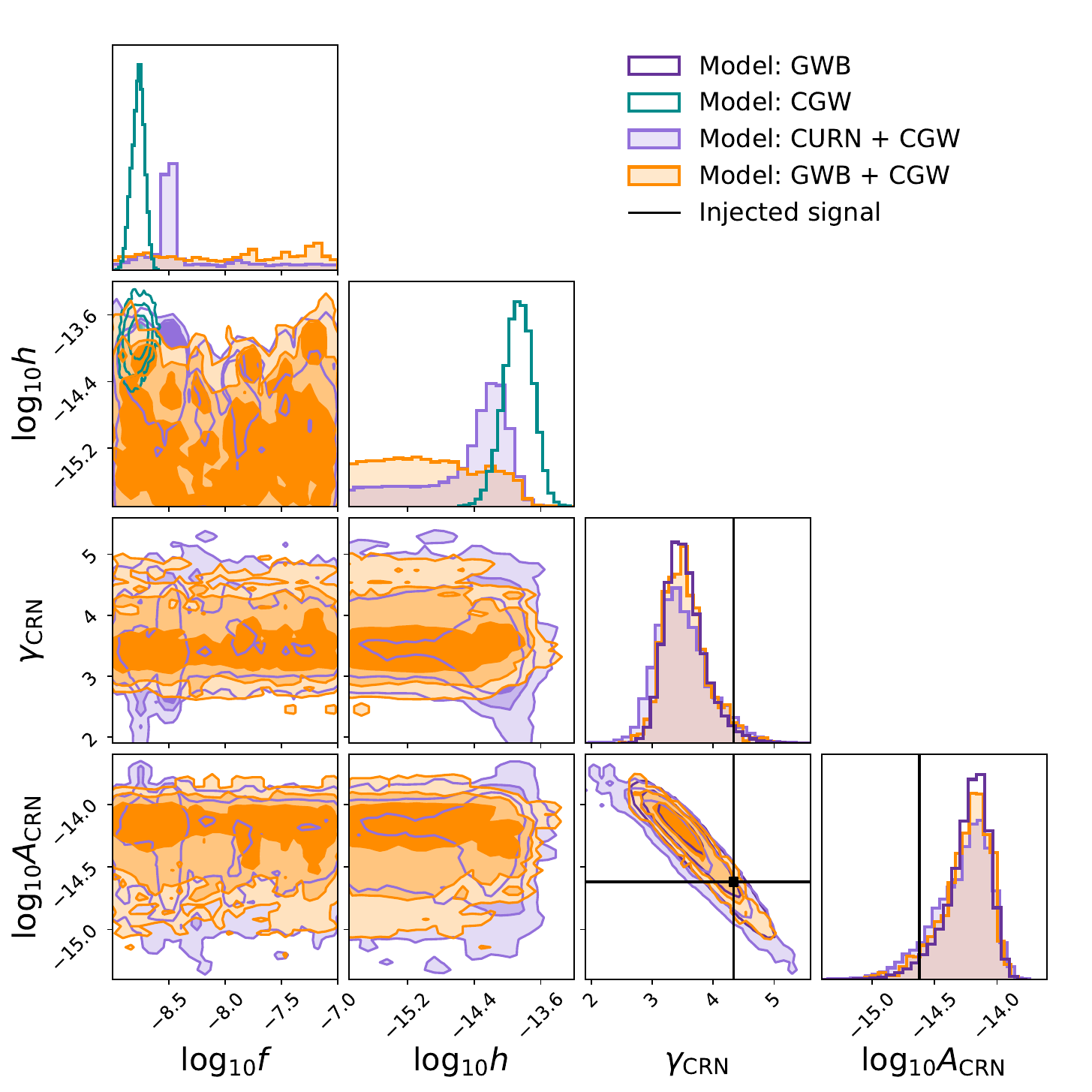}}
\end{minipage}
\begin{minipage}{.5\linewidth}
\centering
{\label{main:b}\includegraphics[scale=.37]{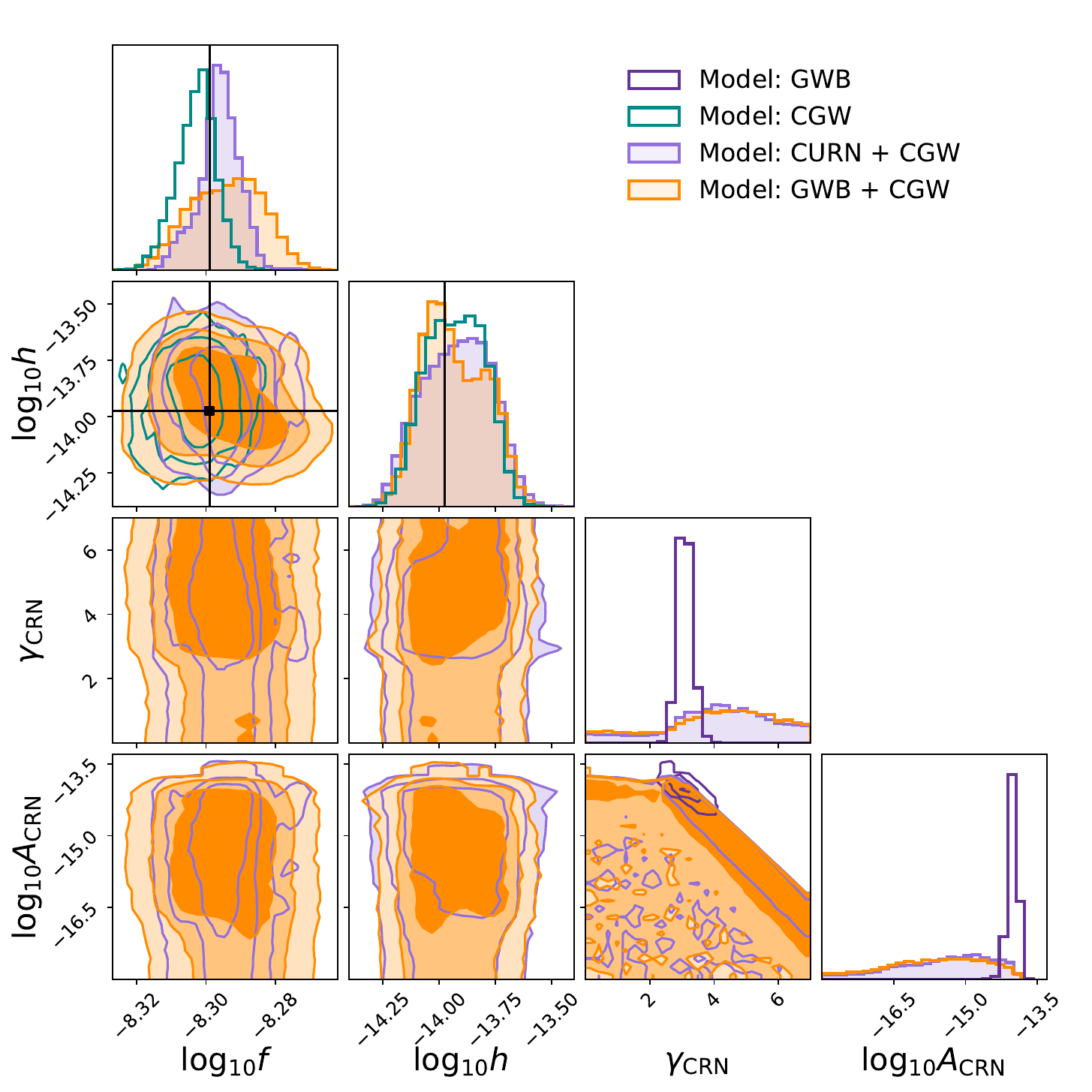}}
\end{minipage}
\caption{Posterior distributions obtained from GWB-only, CGW-only, CURN+CGW and GWB+CGW searches performed on the GWB-only dataset injGWB02 (\textit{left}) and on the CGW-only dataset injCGW5 (\textit{right}).}
\label{fig:gwbcgw_on_data}
\end{figure*}

We now perform a joint CGW + GWB search on the five datasets considered thus far. We perform two flavours of combined searches: (i) common uncorrelated red noise plus CGW (CURN+CGW), and (ii) HD-correlated common red noise plus CGW (GWB+CGW). The CGW is always modelled as the sum of ET and PT, and together with the signal parameters, all the RN and DM parameters are also searched over.

The results of these two searches are reported in Table \ref{gwbcw_vs_gwb} for all datasets, while posterior distributions of selected parameters of the model components are shown in Fig. \ref{fig:gwbcgw_on_data} for datasets injGWB02 and injCGW5. In figure, orange and light purple distributions are obtained respectively from the GWB+CGW and CURN+CGW searches, and results of GWB-only and CGW-only searches are also shown for comparison. To see how searching for both type of signals simultaneously allows to properly characterize the actual content of the data, let us discuss the injGWB and injCGW separately. 

In the injGWB cases, when a common red noise component is added to the model alongside the CGW, the evidence for the latter strongly decreases. This can be seen both from the drop of the Bayes factors in the first three lines of Table \ref{gwbcw_vs_gwb}, and from the posterior distributions of the CGW parameters in the left panel of  Fig. \ref{fig:gwbcgw_on_data}. In Section \ref{GWBasCGW}, when searching for a CGW on a dataset containing only a GWB, we found  $\mathcal{B}^{\rm CGW}_{\rm PSRN}>1000$. This is because the GWB produces a common noise that the model prefers to assign to the CGW signal (common to all pulsars) rather than to the individual pulsars' RN components. 
When a CURN is included in the model, $\mathcal{B}^{\rm CURN+CGW}_{\rm CURN}$ drops significantly but is still slightly larger than unity, suggesting that the presence of CGW component is still preferred. This can also be seen in the posterior distribution of the CGW amplitude shown in purple in the left panel of  Fig. \ref{fig:gwbcgw_on_data} that, while featuring a long tail that extends to the lower end of the prior range, still shows a significant peak around log$A\approx-14.3$. This behaviour is shared by all three injGWB datasets and is due to the fact that the GWB manifests itself primarily as a common red noise process, with a weaker, emerging 'off diagonal' HD-correlated component. The CURN models it as a common red noise process, but is unable to account for the emerging HD correlation, which the model then assigns to the CGW.
Conversely, when a GWB is added to the model, we get $\mathcal{B}^{\rm GWB+CGW}_{\rm GWB}<1$ (first three lines of Table \ref{gwbcw_vs_gwb}), strongly suggesting that the GWB component is now able to describe the full content of the data. This is also clear by looking at the CGW posterior parameters in   the left panel of  Fig. \ref{fig:gwbcgw_on_data}. The orange distributions show that both the frequency and amplitude posteriors are almost perfectly flat, with the latter displaying a cut-off, which we find to be at a similar value for all three realizations. From this we can extract  a 95$\%$ upper limit of the CGW amplitude which is $\approx4.0\times10^{-15}$, $\approx3.8\times10^{-15}$ and $\approx4.6\times10^{-15}$ for datasets injGWB01, injGWB02 and injGWB03 respectively. Note that these upper limits are lower than the GWB level, since the total GW content of the dataset is correctly accounted for by the GWB component. Finally, the recovered source is not localized in the sky, with the posterior distributions of the sky location almost uniform across the sphere, consistent with the prior.

Differently from the CGW parameters, the GWB parameters' posterior distributions are not significantly affected by the addition of the CGW component to the model: the GWB posterior distributions obtained from the GWB+CGW search and with the GWB-only search are well consistent with each other (see left panel of Fig. \ref{fig:gwbcgw_on_data} for the injGWB02 dataset and table \ref{gwbcw_vs_gwb} for all the datasets) confirming again that the GWB component of the model is correctly accounting for all the HD-correlated power. Only a small variation can be noticed in the CURN+CGW recovery of the background parameters: the posterior distributions are slightly wider with a longer low amplitude tail. This is because the CURN cannot absorb any HD correlated power, which is (spuriously) assigned to the CGW component.  

 
Conversely, when the same analysis is applied to the injCGW datasets, the posterior distributions of the CGW parameters are always well constrained, both with the CURN+CGW and with the GWB+CGW search, as shown in the right panel of Fig. \ref{fig:gwbcgw_on_data}. Searching for CGW-only, CURN+CGW or GWB+CGW, in this case, gives consistent results for the CGW parameter, with no significant changes in the posterior distributions' shape and width (see right panel of Fig. \ref{fig:gwbcgw_on_data}). Moreover, GWB+CGW is able to recover the sky location with a precision which is comparable to the one obtained with the CGW-only search and the model comparison strongly favours the GWB+CGW model over the GWB-only model (see table \ref{gwbcw_vs_gwb}). As expected, the GWB posterior distribution is instead completely unconstrained when the model applied in the analysis contains both a common red noise and a CGW: the power that generates the GWB signal recovered with the GWB-only search is now correctly completely absorbed by the CGW component (see Fig. \ref{fig:gwbcgw_on_data}).

\subsubsection{Noise-only datasets}
As a further check, we tested the analysis pipelines on a dataset containing only the noise and no GW signal, neither deterministic nor stochastic. The noise realization is the one contained in all the datasets analysed. In this case, neither a GWB nor a CGW is detected. As for the GWB, we find a value of the Bayes factor $\mathcal{B}^{\rm HD}_{\rm CURN}\sim1.5$ and the posterior distributions of $A$ and $\gamma$ are poorly constrained, with a 95$\%$ upper limit on the GWB amplitude of $\sim 4.47\times 10^{-15}$. As for the CGW, the Bayes factor $\mathcal{B}^{\rm CGW}_{\rm PSRN}$ is $\sim$ 0.5 and the CGW parameters are totally unconstrained, with the amplitude showing a 95$\%$ upper limit of $\sim4.5\times10^{-15}$. Although many more noise realizations should be tested to draw robust statistical conclusions, these tests show on fully realistic datasets that current PTA detection pipelines are unlikely to return a spurious detection in absence of an actual GW signal in the data.

\subsection{Joint GWB+CGW analysis on combined CGW+CGW datasets}\label{combined_searchanddata}

In this section we present the results of simultaneous searches for a GWB and a CGW, applied to datasets injGWB03+CGW5 and injGWB03+CGW20. We remind the reader that both datasets contain the same noise and GWB realisation from dataset injGWB03. 
In each dataset a loud CGW is superimposed to the GWB: in one case the CGW is at 5nHz (injGWB03CGW5), in the second one it is at 20nHz (injGWB03CGW20). This is done in order to see how the joint analysis performs in presence of a slowly evolving source (the 5nHz CGW) and on a fast evolving one (the 20nHz CGW), and how the two of them interact with the GWB.


\subsubsection{Bias induced in the GWB recovery by the presence of a resolvable source in the SMBHBs population}\label{gwb_bias}
As already discussed by \cite{2023ApJ...959....9B} and  \cite{2024A&A...683A.201V}, the presence of a particularly loud source among the systems contributing to the GWB, if not accounted for in the analysis, can significantly bias the recovery of the overall GWB signal. As a preliminary step in our investigation, we therefore perform a GWB only search on the injGWB03CGW5 and injGWB03CGW20 datasets. As expected, the analysis produces a biased recovery of the GWB parameters: the results are shown in the right panels of Fig. \ref{fig:gwbcw_on_11005nHz} and \ref{fig:gwbcw_on_110020nHz}: in both cases, the recovered background is flatter and higher than the injected signal due to the extra power introduced by the loud source. In the two following sections, we demonstrate how this bias can be solved by performing a joint analysis of the GWB and the CGW components.

\subsubsection{Results for the injGWB03+CGW5 dataset}\label{1100cw5}

\begin{figure*}
\begin{minipage}{.5\linewidth}
\centering
{\label{a}\includegraphics[scale=.28]{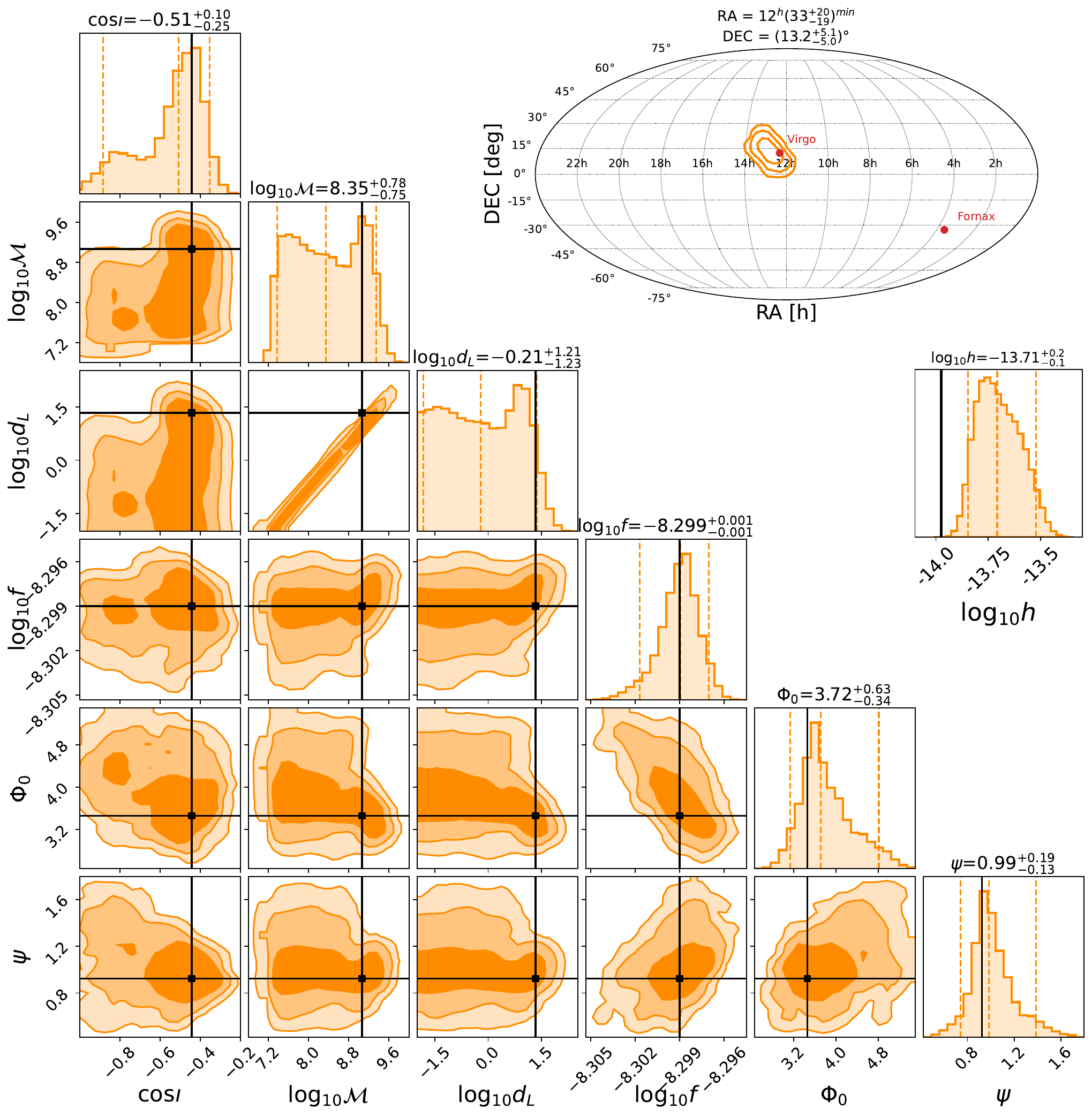}}
\end{minipage}
\begin{minipage}{.5\linewidth}
\centering
{\label{main:b}\includegraphics[scale=.5]{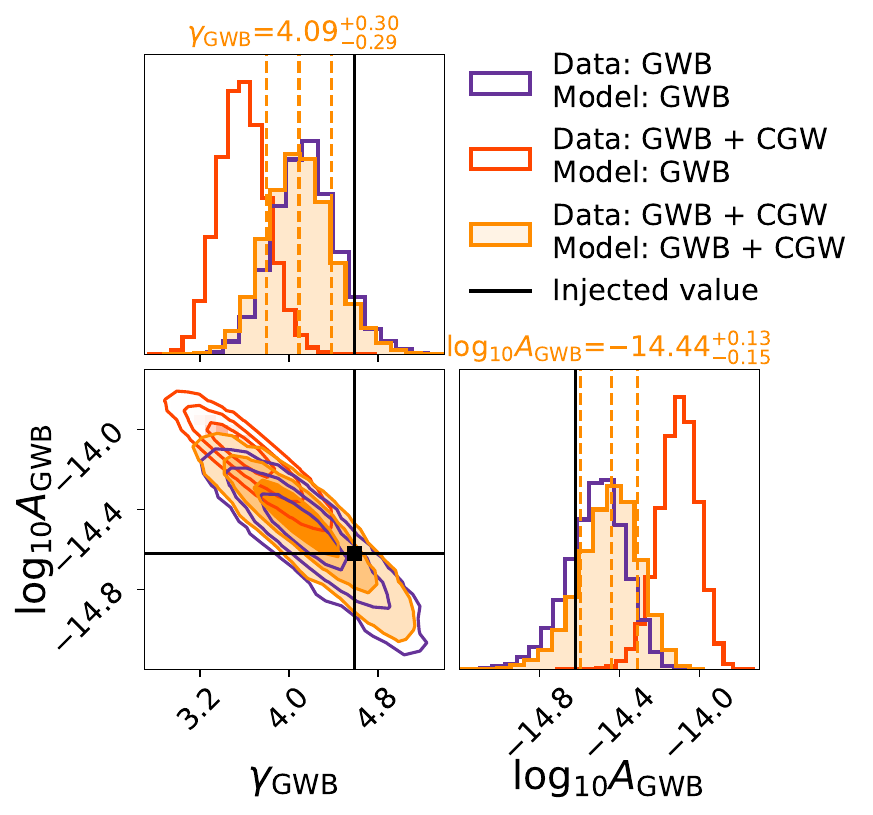}}
\end{minipage}
\caption{CGW {\it left} and GWB {\it right} recovered by the joint search applied on dataset injGWB03CGW5. {\it Left panel:} posterior distributions of the CGW source parameters usint the full ET+PT template. The panel is organized as the ones in Fig.~\ref{fig:etvspt}, with 6 source parameters shown in the triangle plot, the sky localization error shown in the skymap and the recovered signal amplitude $h$ shown in the stand-alone inset. Injected values are marked by the black solid lines. {\it Right panel:}
posterior distributions of the GWB parameters. The GWB recovery from the joint search on the injGWB03+CGW5 dataset is shown in orange. For comparison, the recovery of a GWB only search on the injGWB03CGW5 is shown in red while the recovery of a GWB only search on the GWB-only injGWB03 dataset is shown in purple. Solid black lines represent the injected values. In both panels we mark for each parameter the median and 68\% credible interval of the recovery.
}
\label{fig:gwbcw_on_11005nHz}
\end{figure*}



The results presented here are obtained by sampling the posterior distributions of the CURN+CGW model and then reweighting them to obtain the posteriors of the GWB+CGW model.
The results of the combined search applied to the dataset injGWB03+CGW5 provides good overall results, with the majority of the CGW parameters recovered within 2$\sigma$ of the injected values. 

The posterior distributions of the CGW parameters are shown in the left panel of Fig. \ref{fig:gwbcw_on_11005nHz}, while the recovery of the background parameters is shown in the right panel. The recovery of the CGW parameters is successful: the frequency and the sky localization of the source are well constrained -- the width of the posterior distributions is $\lesssim$10$\%$ of the prior width for all three parameters -- and compatible within 68$\%$ with the injected values: injected frequency is $f = 5.02$nHz, recovered frequency is $f = 5.04^{+0.04}_{-0.04}$nHz; injected sky location is ($12^{\rm h}19^{\rm min}50^{\rm s}$, $12.38^{\circ}$), recovered sky location is ($12^{\rm h} 33^{+20 \rm min}_{-19}$, $10^{+5}_{-5}$$^{\circ}$). However, the 3D source location cannot be properly constrained, since the luminosity distance posterior distribution is  extremely wide due to its strong covariance with the chirp mass. Both posterior distributions are roughly bimodal, with a narrow peak close to the injected value -- in particular for the chirp mass -- and a long, bumpy tail extending to low values of luminosity distance and chirp mass. Note that the posterior of the luminosity distance spans the interval (0.02, 24)Mpc within the 90$\%$ credible region; for such a nearby putative source, ancillary information coming from galaxy catalogues can effectively enable the identification of the source host \citep{2024arXiv240604409P}. The inclination angle, initial phase and polarization angle are well constrained and compatible with the injected values.

Focusing now on the GWB parameters, the posterior distribution in the right panel of Fig. \ref{fig:gwbcw_on_110020nHz} shows how accounting for the resolvable source in the model solves the bias in the GWB parameters' recovery. In fact, the posterior distribution obtained with the combined search is well consistent with the one obtained with the GWB-only search applied to the GWB-only dataset injGWB03: they peak at similar values and have comparable widths, showing that the parameters of the stochastic and of the deterministic components are not significantly correlated and the interplay between the two signals does not affect the search.

\subsubsection{Results for the injGWB03+CGW20 dataset}\label{1100cw20}

\begin{figure*}
\begin{minipage}{.5\linewidth}
\centering
{\label{a}\includegraphics[scale=.28]{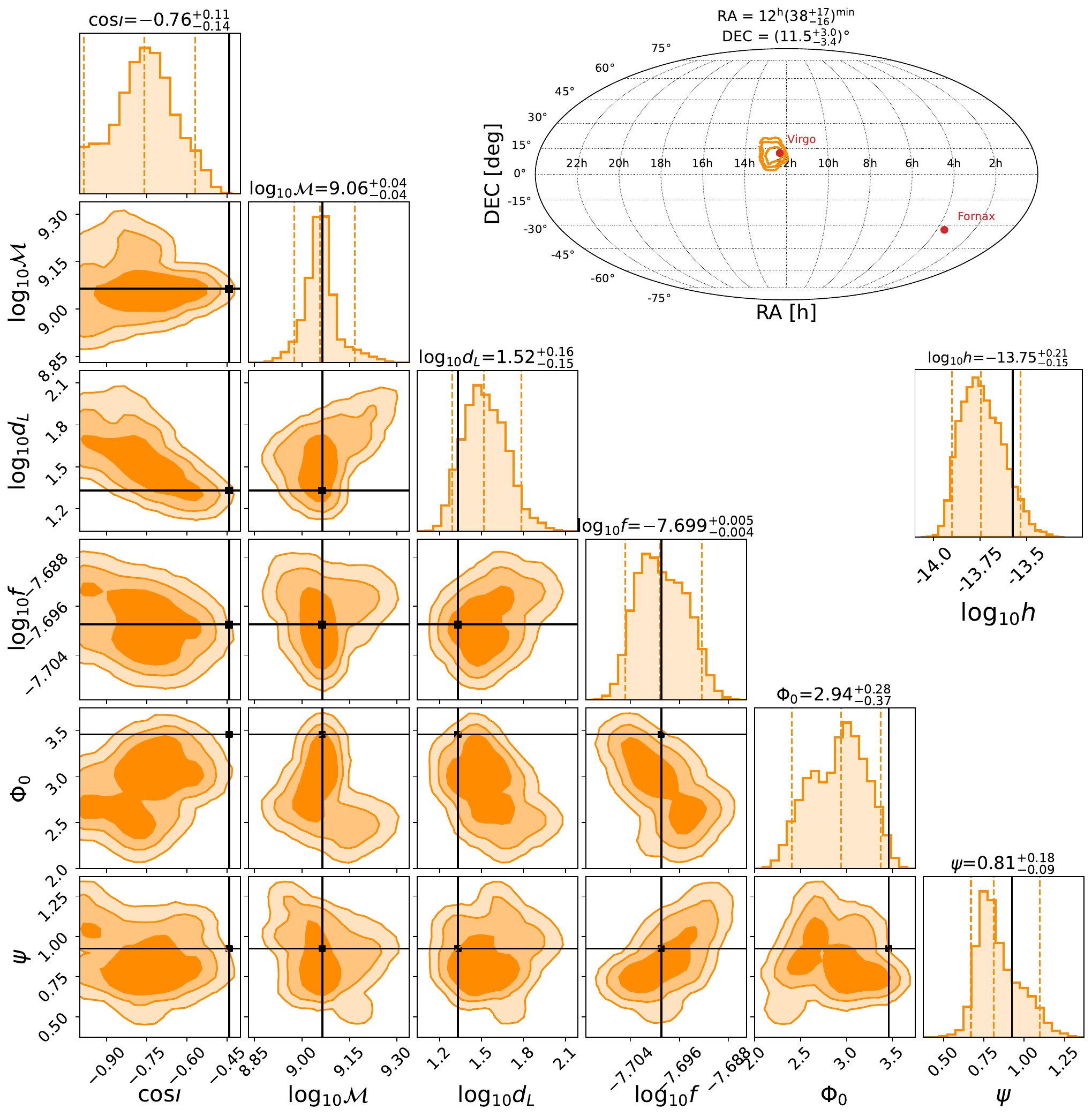}}
\end{minipage}
\begin{minipage}{.5\linewidth}
\centering
{\label{main:b}\includegraphics[scale=.5]{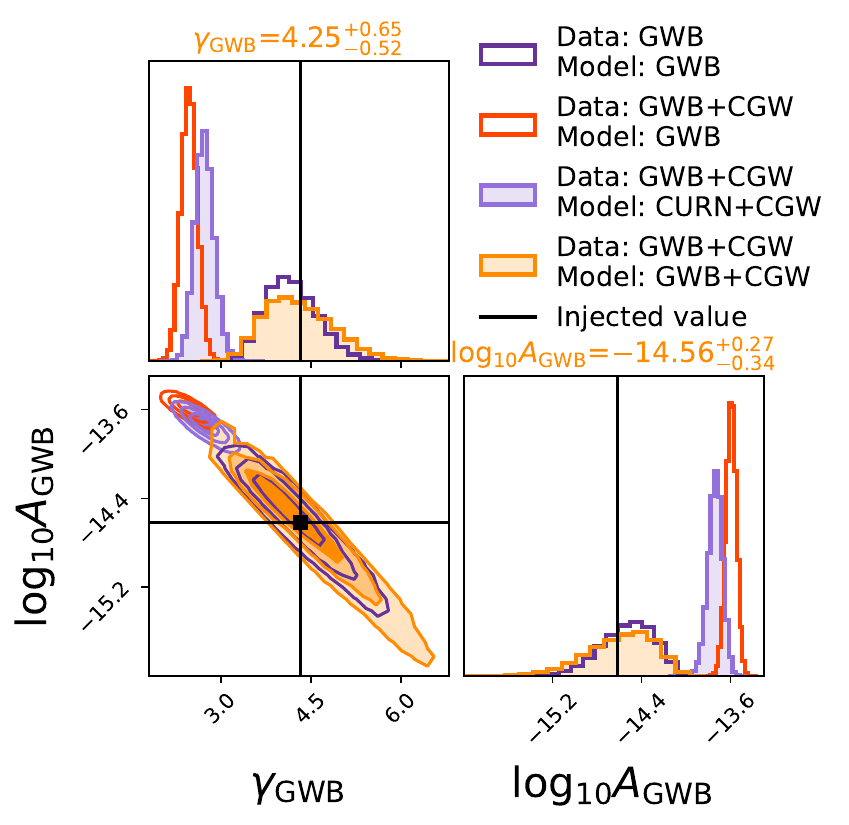}}
\end{minipage}
\caption{Same as Fig.~\ref{fig:gwbcw_on_11005nHz} but for dataset injGWB03+CGW5. The only difference is that in the {\it right panel} we show the joing GWB+CGW analysis either modelling the GWB as a CURN and then using the reweighting technique (filled purple) or by directly modelling the HD correlations in the GWB search (filled orange). All other distributions and markers have the same color style and meaning as in Fig.~\ref{fig:gwbcw_on_11005nHz}.
}
\label{fig:gwbcw_on_110020nHz}
\end{figure*}


\begin{figure*}
   \centering
   \includegraphics[width=1\textwidth]{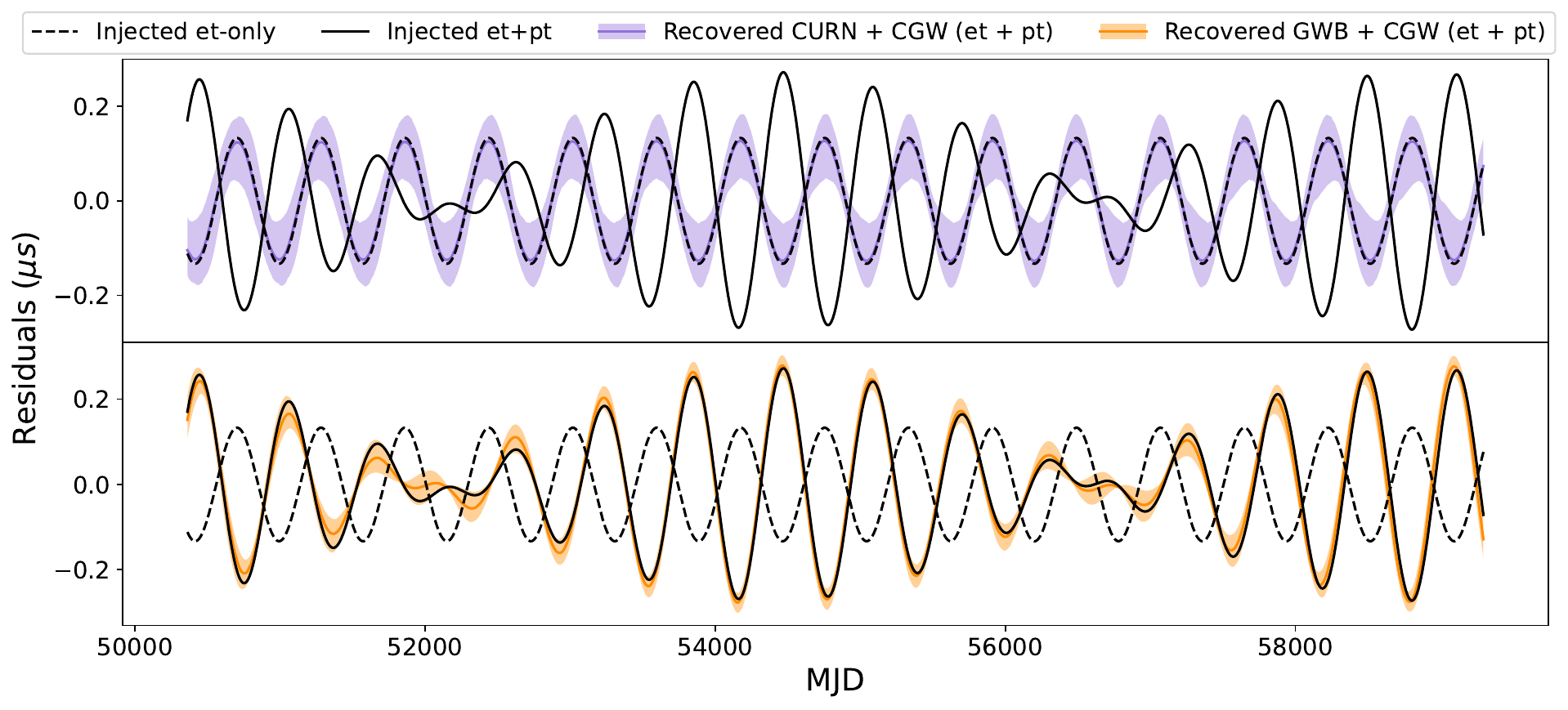}
   \caption{Median and 90$\%$ credible region of the recovered CGW signal for pulsar PSR J1713+4707 from the joint analysis of dataset injGWB03+CGW5. The {\it top panel} shows the result obtained with a CURN+CGW (ET+PT) search and {\it bottom panel} shows the outcome of a GWB+CGW (ET+PT) search. Solid ad dashed line are respectively the total and the ET-only signal injected. In the first case, the CGW model neglects the pulsar term, resulting in a biased CGW parameters recovery (detail explanation in the main text).}
   \label{fig:gwbcw_res}%
\end{figure*}

The joint search applied to the injGWB03CGW20 dataset unveils some new interesting behaviour of the recovery pipeline. In fact, in this case, when a joint CURN+CGW (ET+PT) search is performed and the GWB is obtained with the reweighting technique, the estimate of the GWB parameters is completely off, as demonstrated by the filled purple distributions shown in the right panel of 
Fig.~\ref{fig:gwbcw_on_110020nHz}. The result is indeed quite similar to a GWB search only (shown in red), meaning that a significant fraction of the CGW power is absorbed in the common background component and the CURN+CGW search followed by reweighting fails in this case, producing strong biases in the results.


The reason of this failure is the following: as shown in Section \ref{sec:etvspt}, due to the significant difference in the 20nHz CGW between the Earth and the pulsar term GW frequency, in order to correctly account for both terms with the CGW model, the majority of the 50 PT parameters (i.e. 25 pulsar distances and 25 individual PT GW phases) have to be constrained. This is a difficult task for the sampler, which in the CURN+CGW case converges to an easier, alternative solution: the pulsar terms are all absorbed by the common uncorrelated red noise component, which is possible because they are indeed a common sinusoidal process that is uncorrelated, since the pulsars are all at different sky locations and distances from Earth. This justifies the much higher and flatter recovered background shown in the right panel of Fig. \ref{fig:gwbcw_on_110020nHz}: the recovered amplitude is $1.83^{+0.28}_{-0.26}\times10^{-14}$, while the injected value is $2.4\times10^{-15}$. 

This interpretation is confirmed by the recovered CGW parameters (not shown). Most of them are biased and in particular the chirp mass, which determines the frequency evolution of the signal, is much lower than the injected value: the obtained value is $6. 9^{+18.0}_{-5.0}\times10^{6}{\rm M}_{\odot}$, which corresponds to a 95$\%$ upper limit on the frequency derivative of $4.2\times10^{-6}$nHz$\cdot$yr$^{-1}$, $\sim$300 times smaller than the $\dot{f}$ of the injected SMBHB. This is consistent with the fact that the values of the CGW parameters are such that the recovered time domain signal matches the injected Earth term and completely neglects the pulsar term contribution. This is confirmed by the shape of the reconstructed CGW shown in  Fig.\ref{fig:gwbcw_res}. In the top panel we can see that in the CURN+CGW search the recovered CGW waveform for pulsar PSR J1713+0747 reproduces exactly the injected ET contribution. In fact, being HD-correlated, in presence of a CGW component in the model, the Earth term is not absorbed by the CURN component. Indeed, the recovered background is slightly lower than the one obtained by a GWB-only search on the same dataset (shown by the red distributions in the right panel of Fig. \ref{fig:gwbcw_on_110020nHz}), because in that case both the Earth and pulsar terms are absorbed by the background component. 

We stress that this bias indeed corresponds to the maximum of the posterior distribution. This can be explained by the fact that the CURN model does not exactly match the injected common red noise, which is HD correlated, and therefore it prefers to absorb the pulsar term, since it is much cheaper to describe it as a two-parameter CURN signal, rather than to determine the 2$\times$25 parameters describing each individual sinusoidal pulsar term.

In this dataset, the injected signal components are  properly recovered only by a direct GWB+CGW (ET+PT) search, which is more computationally expensive due to the off-diagonal HD terms that must be computed when evaluating the correlation matrix in Eq.~\eqref{LIke}. In fact, in a GWB+CGW search the pulsar term is not absorbed by the HD-correlated common red noise component and both signals are properly reconstructed. In the right panel of Fig.\ref{fig:gwbcw_on_110020nHz} the GWB recovery obtained by this search is shown in filled purple: it agrees very well with that obtained with the GWB-only template applied to the GWB-only dataset (empty purple), although it is slightly larger, which can be explained by the fact that the GWB+CGW model contains 58 more parameters describing the single source. A tiny correlation with this high number of parameters can easily explain the less precise recovery of the background parameters.

The correct recovery of the GWB parameters is a direct consequence of the correct recovery of the CGW parameters. In the left panel of Fig. \ref{fig:gwbcw_on_110020nHz}, the posterior distributions of the 8 source parameters are shown and in Fig. \ref{fig:gwbcw_res} the median and 90$\%$ credible region of the recovered CGW signal is shown for pulsar PSR J1713+4707. From the recovery, it can be seen how the model is able to account for the frequency evolution of the signal: the frequency recovered is $19.99^{+0.21}_{0.18}$nHz (injected is 20nHz) and the chirp mass $1.14^{+0.11}_{0.10}\times10^{9}M_{\odot}$ (injected is $1.16\times 10^{9}M_{\odot}$). This allows the luminosity distance to be much better constrained than in the low frequency case, with a median and 68$\%$ credible interval of $32.7^{+15.1}_{-9.3}$Mpc. Since the injected value is 21.6Mpc, the value obtained is a mild overestimate, which is probably due to the anti correlation with the inclination angle and the initial phase, which are both slightly underestimated. The 2D sky location recovery is also improved with respect to the 5nHz case, resulting in a declination angle of $12^{\rm h}38^{+17\rm min}_{-16}$ and a right ascension of $11.5^{+3.0}_{-3.4}$$^{\circ}$.

\section{Discussion and conclusions}\label{sec:discussion}

In this paper, we have carried out the first comprehensive study of the interplay between stochastic GWBs and deterministic CGWs in realistic PTA datasets, assessing the performance of currently used analysis pipelines. To this end we constructed mock datasets featuring all the complexity of EPTA DR2full \citep{2023A&A...678A..48E}; we simulated up to 24.8 years of unevenly sampled, multi frequency ToAs for the 25 pulsars of DR2full and injected WN, RN and DM consistent with the maximum likelihood values found in the DR2full single pulsar analysis \citep{2023}. The injected GW signals were also as realistic as possible, tailored to mimic the findings of the recent EPTA analyses \citep{EPTA}. A GWB with nominal amplitude of $A_{\rm GW}=2.4\times 10^{-15}$ was injected as the incoherent sum of individual waves coming from observationally based cosmic populations of SMBHBs \citep{2024A&A...685A..94E}, while the deterministic CGW was modelled according to the candidate signal identified in the analysis of EPTA DR2new \citep{CGW_EPTA}. In all cases, SMBHBs were assumed to be circular and both the ET and PT of the GW induced time delays were injected. These GW signals were searched with the standard models implemented in the \texttt{ENTERPRISE} analysis suite. The GWB was therefore modelled as a common red process characterized by a single power law spectrum, while the model CGW matched the one used for injecting the signal (i.e., circular, GW driven SMBHB), either including only ET or ET+PT. With this set-up we carried a number of investigations, whose results are summarized in the following.

We first compared the performance of ET-only and ET+PT templates on two datasets containing only a CGW either at 5nHz (injCGW5) or at 20nHz (injCGW20). In both cases, the ET+PT template was able to fully recover the signal, providing an unbiased estimate of all its parameters, including chirp mass, distance and sky location. Things were not so straightforward when the ET only template was employed. In this case, the 5nHz CGW dataset resulted in a biased recovery of the frequency and the sky location, whereas in the 20nHz case, the recovery was unbiased. In fact, in the latter case Earth and pulsar term can be easily disentangled and the ET-only template correctly matches the ET component of the signal. Conversely, in the 5 nHz case, ET and PT fall in the same frequency bin and the ET template tries to fit a mixture of the two, leading to a bias in the signal reconstruction. In both cases, the ET-only model is not sensitive to the chirp mass and therefore to the luminosity distance, preventing 3D source localization.
We also stress that in both cases, modelling ET only leaves behind some unmodelled residual (e.g. the full PT in the 20 nHz case), which will likely be absorbed by the GWB template in a joint search, leading to a biased recovery of the background parameters. These are the reasons why the complete CGW template (ET+PT) must be preferred to the ET-only, even if this means having 2 times the number of pulsars extra parameters (50 in our case) and therefore much larger computational costs for the sampling.

We therefore adopted the full ET+PT template for the CGW and turned to the investigation of datasets in which only one of the two types of signal was injected (either a GWB or a CGW) and the analyses were performed separately with one model at the time (eiter a GWB or a CGW). We analysed 5 datasets: 3 realizations of a realistic GWB (dataset injGWB01,injGWB02, injGWB03) and 2 realizations of single resolvable sources (datasets injCGW5, injCGW20). All datasets contained the same noise realization. These searches showed that a GWB can easily be misinterpreted as a CGW and vice versa. In fact, all sources returned evidence in favour of either GW signal and well constrained recovery of the model parameters. However, when a combined search is performed -- meaning that the model used for the recovery contains both an HD-correlated GWB and a CGW -- this degeneracy is broken: when only a GWB is present in the data, the posterior distributions of the CGW parameters obtained with the combined search are unconstrained and the CGW amplitude shows an upper limit; on the contrary, when a CGW is the only signal injected, the combined search gives unconstrained posterior distributions for the GWB parameters. No CGW evidence is absorbed by the HD-correlated background if the single source is present in the data. Therefore, the combined search is able to correctly recover the signal that is actually present in the data. Focusing on GWB-only datasets, we highlight that the CGW evidence is totally absorbed when an HD-correlated GWB is present in the model, while it does not completely disappear if only a CURN component is modelled. In this case, even if poorly constrained, the CGW parameters posterior distributions still peak at specific values. Therefore, modelling the common red noise as HD-correlated is necessary to totally absorb the spurious CGW. These results are similar to the ones obtained with the real EPTA DR2new data \citep{CGW_EPTA}: CURN + CGW searches peaking at a CGW amplitude of $\sim10^{-14}$, and peak disappearing once the background is modelled as HD-correlated and the Bayes factor $\mathcal{B}^{\rm GWB+CGW}_{\rm GWB}$ being of order unity. Our simulations show that this behaviour arises when a CGW is indeed not present in the data, strengthening 
  the conclusion reached in 
the EPTA DR2 analysis.

We then analysed datasets containing both a stochastic GWB and a CGW, either at 5 nHz (injGWB03+CGW5) or at 20 nHz (injGWB03+CGW20). The low frequency CGW is an almost monochromatic source, while the high frequency one is drifting and the signal frequency significantly changes between the  PT and the ET. Performing a combined search on these types of datasets is particularly important because a single resolvable source, if not properly taken into account in the search, will induce a bias in the GWB parameters recovery \citep{2023ApJ...959....9B,2024A&A...683A.201V}.
The results of the combined searches indicate that the bias in the GWB recovery induced by the presence of the single source is resolved when the CGW is correctly accounted for by the deterministic CGW model, so long as it contains both ET and PT. This result is obtained with both datasets, injGWB03CGW5 and injGWB03CGW20. However, in the injGWB03CGW5 case, the same unbiased result can be obtained with two methods: by performing a CURN+CGW search and then reweighing to obtain the HD+CGW posterior distribution, and by directly sampling the HD+CGW posterior distribution. The first method is computationally advantageous, as the CURN model is considerably faster to compute than the HD one. In the injGWB03CGW20 case, however, the first method (reweighting) is unable to recover the unbiased posterior distribution, as the CGW template solely accounts for the ET contribution, while the PT is more easily absorbed by the CURN. The posterior distribution on the CURN parameters is therefore strongly biased with respect to the GWB injection. Performing a direct sampling of the HD+CGW model, on the contrary, gives an unbiased recovery of the GWB parameters: the posterior distributions of the CURN and HD parameters are consequently disjoint, which makes the reweighing technique extremely inefficient. Therefore, while computationally much more expensive, the only method that is able to converge to unbiased results is the direct HD+CGW search. As for the GWB, the CGW recovery is successful, with most of the parameters constrained and compatible with the injection.

Given the recent PTA results indicating the presence of a gravitational signal in the data, it is crucial to be able to distinguish between the two types of signal that can be produced by SMBHBs. This work represents a first thorough study of this problem. Despite the limited parameter exploration, which is constrained by the high computational cost of the analyses, the investigation decisively points towards the importance of performing combined searches. Indeed, in the cases analysed, combined searches were able to both identify which signal was present in the data, without misinterpretation, and to guarantee an unbiased recovery of the GWB parameters in the presence of a loud individual source in the SMBHB population. Looking ahead, it will be important to extend this analysis to richer and more sensitive datasets, investigating the possibility of disentangling multiple searches, also in the presence of eccentricity or environmental coupling. In fact, the high amplitude of the currently observed signal implies that the future SKA might be able to resolve several tens of individual SMBHBs at $S/N>5$ \citep{2024arXiv240712078T}. Being able to successfully identify and disentangle those signals will have invaluable payouts for the understanding of SMBHB evolution and dynamics in the context of multimessenger astrophysics. Given the complexity and high dimensionality of these sources, it is of paramount importance to design more efficient search techniques \citep{2024arXiv240616331B,2024arXiv240711135F,2024arXiv240508857V}.

\begin{acknowledgements}
       I.F. thanks Stanislav Babak for his hospitality and mentoring at the APC, where this investigation was initiated.
       We thank the B-Massive group at Milano-Bicocca University for useful discussions and comments. A.C., A.S., G.M.S. acknowledge the financial support provided under the European Union’s H2020 ERC Consolidator Grant ``Binary Massive Black Hole Astrophysics'' (B Massive, Grant Agreement: 818691). A.C. acknowledges financial support provided under the European Union's Horizon Europe ERC Starting Grant "A Gamma-ray Infrastructure to Advance Gravitational Wave Astrophysics" (GIGA).
\end{acknowledgements}

\bibliographystyle{aa}
\bibliography{bibliography}

\begin{thebibliography}{67}
\expandafter\ifx\csname natexlab\endcsname\relax\def\natexlab#1{#1}\fi

\bibitem[{{Afzal} {et~al.}(2023){Afzal}, {Agazie}, {Anumarlapudi}, {Archibald}, {Arzoumanian}, {Baker}, {B{\'e}csy}, {Blanco-Pillado}, {Blecha}, {Boddy}, {Brazier}, {Brook}, {Burke-Spolaor}, {Burnette}, {Case}, {Charisi}, {Chatterjee}, {Chatziioannou}, {Cheeseboro}, {Chen}, {Cohen}, {Cordes}, {Cornish}, {Crawford}, {Cromartie}, {Crowter}, {Cutler}, {Decesar}, {Degan}, {Demorest}, {Deng}, {Dolch}, {Drachler}, {von Eckardstein}, {Ferrara}, {Fiore}, {Fonseca}, {Freedman}, {Garver-Daniels}, {Gentile}, {Gersbach}, {Glaser}, {Good}, {Guertin}, {G{\"u}ltekin}, {Hazboun}, {Hourihane}, {Islo}, {Jennings}, {Johnson}, {Jones}, {Kaiser}, {Kaplan}, {Kelley}, {Kerr}, {Key}, {Laal}, {Lam}, {Lamb}, {Lazio}, {Lee}, {Lewandowska}, {Lino Dos Santos}, {Littenberg}, {Liu}, {Lorimer}, {Luo}, {Lynch}, {Ma}, {Madison}, {McEwen}, {McKee}, {McLaughlin}, {McMann}, {Meyers}, {Meyers}, {Mingarelli}, {Mitridate}, {Nay}, {Natarajan}, {Ng}, {Nice}, {Ocker}, {Olum}, {Pennucci}, {Perera}, {Petrov}, {Pol}, {Radovan}, {Ransom}, {Ray}, {Romano},
  {Sardesai}, {Schmiedekamp}, {Schmiedekamp}, {Schmitz}, {Schr{\"o}der}, {Schult}, {Shapiro-Albert}, {Siemens}, {Simon}, {Siwek}, {Stairs}, {Stinebring}, {Stovall}, {Stratmann}, {Sun}, {Susobhanan}, {Swiggum}, {Taylor}, {Taylor}, {Trickle}, {Turner}, {Unal}, {Vallisneri}, {Verma}, {Vigeland}, {Wahl}, {Wang}, {Witt}, {Wright}, {Young}, {Zurek}, \& {Nanograv Collaboration}}]{2023ApJ...951L..11A}
{Afzal}, A., {Agazie}, G., {Anumarlapudi}, A., {et~al.} 2023, \apjl, 951, L11

\bibitem[{{Agazie} {et~al.}(2023{\natexlab{a}}){Agazie}, {Anumarlapudi}, {Archibald}, {Arzoumanian}, {Baker}, {B{\'e}csy}, {Blecha}, {Brazier}, {Brook}, {Burke-Spolaor}, {Case}, {Casey-Clyde}, {Charisi}, {Chatterjee}, {Cohen}, {Cordes}, {Cornish}, {Crawford}, {Cromartie}, {Crowter}, {Decesar}, {Demorest}, {Digman}, {Dolch}, {Drachler}, {Ferrara}, {Fiore}, {Fonseca}, {Freedman}, {Garver-Daniels}, {Gentile}, {Glaser}, {Good}, {G{\"u}ltekin}, {Hazboun}, {Hourihane}, {Jennings}, {Johnson}, {Jones}, {Kaiser}, {Kaplan}, {Kelley}, {Kerr}, {Key}, {Laal}, {Lam}, {Lamb}, {Lazio}, {Lewandowska}, {Liu}, {Lorimer}, {Luo}, {Lynch}, {Ma}, {Madison}, {McEwen}, {McKee}, {McLaughlin}, {McMann}, {Meyers}, {Meyers}, {Mingarelli}, {Mitridate}, {Ng}, {Nice}, {Ocker}, {Olum}, {Pennucci}, {Perera}, {Petrov}, {Pol}, {Radovan}, {Ransom}, {Ray}, {Romano}, {Sardesai}, {Schmiedekamp}, {Schmiedekamp}, {Schmitz}, {Shapiro-Albert}, {Siemens}, {Simon}, {Siwek}, {Stairs}, {Stinebring}, {Stovall}, {Susobhanan}, {Swiggum}, {Taylor}, {Taylor},
  {Turner}, {Unal}, {Vallisneri}, {van Haasteren}, {Vigeland}, {Wahl}, {Witt}, {Young}, \& {Nanograv Collaboration}}]{2023ApJ...951L..50A}
{Agazie}, G., {Anumarlapudi}, A., {Archibald}, A.~M., {et~al.} 2023{\natexlab{a}}, \apjl, 951, L50

\bibitem[{{Agazie} {et~al.}(2023{\natexlab{b}}){Agazie}, {Anumarlapudi}, {Archibald}, {Arzoumanian}, {Baker}, {B{\'e}csy}, {Blecha}, {Brazier}, {Brook}, {Burke-Spolaor}, {Casey-Clyde}, {Charisi}, {Chatterjee}, {Cohen}, {Cordes}, {Cornish}, {Crawford}, {Cromartie}, {Crowter}, {DeCesar}, {Demorest}, {Dolch}, {Drachler}, {Ferrara}, {Fiore}, {Fonseca}, {Freedman}, {Gardiner}, {Garver-Daniels}, {Gentile}, {Glaser}, {Good}, {G{\"u}ltekin}, {Hazboun}, {Jennings}, {Johnson}, {Jones}, {Kaiser}, {Kaplan}, {Kelley}, {Kerr}, {Key}, {Laal}, {Lam}, {Lamb}, {Lazio}, {Lewandowska}, {Liu}, {Lorimer}, {Luo}, {Lynch}, {Ma}, {Madison}, {McEwen}, {McKee}, {McLaughlin}, {McMann}, {Meyers}, {Mingarelli}, {Mitridate}, {Ng}, {Nice}, {Ocker}, {Olum}, {Pennucci}, {Perera}, {Pol}, {Radovan}, {Ransom}, {Ray}, {Romano}, {Sardesai}, {Schmiedekamp}, {Schmiedekamp}, {Schmitz}, {Schult}, {Shapiro-Albert}, {Siemens}, {Simon}, {Siwek}, {Stairs}, {Stinebring}, {Stovall}, {Susobhanan}, {Swiggum}, {Taylor}, {Turner}, {Unal}, {Vallisneri},
  {Vigeland}, {Wahl}, {Witt}, \& {Young}}]{NANOGrav}
{Agazie}, G., {Anumarlapudi}, A., {Archibald}, A.~M., {et~al.} 2023{\natexlab{b}}, arXiv e-prints, arXiv:2306.16221

\bibitem[{{Ali-Ha{\"\i}moud} {et~al.}(2020){Ali-Ha{\"\i}moud}, {Smith}, \& {Mingarelli}}]{2020PhRvD.102l2005A}
{Ali-Ha{\"\i}moud}, Y., {Smith}, T.~L., \& {Mingarelli}, C. M.~F. 2020, \prd, 102, 122005

\bibitem[{{Allen}(2023)}]{2023PhRvD.107d3018A}
{Allen}, B. 2023, \prd, 107, 043018

\bibitem[{Allen(2023)}]{Allen_2023}
Allen, B. 2023, Physical Review D, 107

\bibitem[{{Antoniadis} {et~al.}(2023){Antoniadis}, {Arumugam}, {Arumugam}, {Babak}, {Bagchi}, {Bak Nielsen}, {Bassa}, {Bathula}, {Berthereau}, {Bonetti}, {Bortolas}, {Brook}, {Burgay}, {Caballero}, {Chalumeau}, {Champion}, {Chanlaridis}, {Chen}, {Cognard}, {Dandapat}, {Deb}, {Desai}, {Desvignes}, {Dhanda-Batra}, {Dwivedi}, {Falxa}, {Ferranti}, {Ferdman}, {Franchini}, {Gair}, {Goncharov}, {Gopakumar}, {Graikou}, {Grie{\ss}meier}, {Guillemot}, {Guo}, {Gupta}, {Hisano}, {Hu}, {Iraci}, {Izquierdo-Villalba}, {Jang}, {Jawor}, {Janssen}, {Jessner}, {Joshi}, {Kareem}, {Karuppusamy}, {Keane}, {Keith}, {Kharbanda}, {Kikunaga}, {Kolhe}, {Kramer}, {Krishnakumar}, {Lackeos}, {Lee}, {Liu}, {Liu}, {Lyne}, {McKee}, {Maan}, {Main}, {Manzini}, {Mickaliger}, {Nitu}, {Nobleson}, {Paladi}, {Parthasarathy}, {Perera}, {Perrodin}, {Petiteau}, {Porayko}, {Possenti}, {Prabu}, {Quelquejay Leclere}, {Rana}, {Samajdar}, {Sanidas}, {Sesana}, {Shaifullah}, {Singha}, {Speri}, {Spiewak}, {Srivastava}, {Stappers}, {Surnis}, {Susarla},
  {Susobhanan}, {Takahashi}, {Tarafdar}, {Theureau}, {Tiburzi}, {van der Wateren}, {Vecchio}, {Venkatraman Krishnan}, {Verbiest}, {Wang}, {Wang}, \& {Wu}}]{CGW_EPTA}
{Antoniadis}, J., {Arumugam}, P., {Arumugam}, S., {et~al.} 2023, arXiv e-prints, arXiv:2306.16226

\bibitem[{{Arzoumanian} {et~al.}(2016){Arzoumanian}, {Brazier}, {Burke-Spolaor}, {Chamberlin}, {Chatterjee}, {Christy}, {Cordes}, {Cornish}, {Crowter}, {Demorest}, {Deng}, {Dolch}, {Ellis}, {Ferdman}, {Fonseca}, {Garver-Daniels}, {Gonzalez}, {Jenet}, {Jones}, {Jones}, {Kaspi}, {Koop}, {Lam}, {Lazio}, {Levin}, {Lommen}, {Lorimer}, {Luo}, {Lynch}, {Madison}, {McLaughlin}, {McWilliams}, {Mingarelli}, {Nice}, {Palliyaguru}, {Pennucci}, {Ransom}, {Sampson}, {Sanidas}, {Sesana}, {Siemens}, {Simon}, {Stairs}, {Stinebring}, {Stovall}, {Swiggum}, {Taylor}, {Vallisneri}, {van Haasteren}, {Wang}, {Zhu}, \& {NANOGrav Collaboration}}]{2016ApJ...821...13A}
{Arzoumanian}, Z., {Brazier}, A., {Burke-Spolaor}, S., {et~al.} 2016, \apj, 821, 13

\bibitem[{Babak \& Sesana(2012)}]{Babak_2012}
Babak, S. \& Sesana, A. 2012, Physical Review D, 85

\bibitem[{{B{\'e}csy}(2024)}]{2024arXiv240616331B}
{B{\'e}csy}, B. 2024, arXiv e-prints, arXiv:2406.16331

\bibitem[{{B{\'e}csy} \& {Cornish}(2020)}]{2020CQGra..37m5011B}
{B{\'e}csy}, B. \& {Cornish}, N.~J. 2020, Classical and Quantum Gravity, 37, 135011

\bibitem[{{B{\'e}csy} {et~al.}(2022){B{\'e}csy}, {Cornish}, \& {Kelley}}]{2022ApJ...941..119B}
{B{\'e}csy}, B., {Cornish}, N.~J., \& {Kelley}, L.~Z. 2022, \apj, 941, 119

\bibitem[{{B{\'e}csy} {et~al.}(2023){B{\'e}csy}, {Cornish}, {Meyers}, {Kelley}, {Agazie}, {Anumarlapudi}, {Archibald}, {Arzoumanian}, {Baker}, {Blecha}, {Brazier}, {Brook}, {Burke-Spolaor}, {Casey-Clyde}, {Charisi}, {Chatterjee}, {Chatziioannou}, {Cohen}, {Cordes}, {Crawford}, {Cromartie}, {Crowter}, {DeCesar}, {Demorest}, {Dolch}, {Ferrara}, {Fiore}, {Fonseca}, {Freedman}, {Garver-Daniels}, {Gentile}, {Glaser}, {Good}, {G{\"u}ltekin}, {Hazboun}, {Hourihane}, {Jennings}, {Johnson}, {Jones}, {Kaiser}, {Kaplan}, {Kerr}, {Key}, {Laal}, {Lam}, {Lamb}, {W. Lazio}, {Lewandowska}, {Littenberg}, {Liu}, {Lorimer}, {Luo}, {Lynch}, {Ma}, {Madison}, {McEwen}, {McKee}, {McLaughlin}, {McMann}, {Meyers}, {Mingarelli}, {Mitridate}, {Ng}, {Nice}, {Ocker}, {Olum}, {Pennucci}, {Perera}, {Pol}, {Radovan}, {Ransom}, {Ray}, {Romano}, {Sardesai}, {Schmiedekamp}, {Schmiedekamp}, {Schmitz}, {Shapiro-Albert}, {Siemens}, {Simon}, {Siwek}, {Sosa Fiscella}, {Stairs}, {Stinebring}, {Stovall}, {Susobhanan}, {Swiggum}, {Taylor}, {Turner},
  {Unal}, {Vallisneri}, {van Haasteren}, {Vigeland}, {Wahl}, {Witt}, \& {Young}}]{2023ApJ...959....9B}
{B{\'e}csy}, B., {Cornish}, N.~J., {Meyers}, P.~M., {et~al.} 2023, \apj, 959, 9

\bibitem[{{Burke-Spolaor}(2013)}]{2013CQGra..30v4013B}
{Burke-Spolaor}, S. 2013, Classical and Quantum Gravity, 30, 224013

\bibitem[{{Chalumeau} {et~al.}(2022){Chalumeau}, {Babak}, {Petiteau}, {Chen}, {Samajdar}, {Caballero}, {Theureau}, {Guillemot}, {Desvignes}, {Parthasarathy}, {Liu}, {Shaifullah}, {Hu}, {van der Wateren}, {Antoniadis}, {Bak Nielsen}, {Bassa}, {Berthereau}, {Burgay}, {Champion}, {Cognard}, {Falxa}, {Ferdman}, {Freire}, {Gair}, {Graikou}, {Guo}, {Jang}, {Janssen}, {Karuppusamy}, {Keith}, {Kramer}, {Lee}, {Liu}, {Lyne}, {Main}, {McKee}, {Mickaliger}, {Perera}, {Perrodin}, {Porayko}, {Possenti}, {Sanidas}, {Sesana}, {Speri}, {Stappers}, {Tiburzi}, {Vecchio}, {Verbiest}, {Wang}, {Wang}, \& {Xu}}]{2022MNRAS.509.5538C}
{Chalumeau}, A., {Babak}, S., {Petiteau}, A., {et~al.} 2022, \mnras, 509, 5538

\bibitem[{{Charisi} {et~al.}(2022){Charisi}, {Taylor}, {Runnoe}, {Bogdanovic}, \& {Trump}}]{2022MNRAS.510.5929C}
{Charisi}, M., {Taylor}, S.~R., {Runnoe}, J., {Bogdanovic}, T., \& {Trump}, J.~R. 2022, \mnras, 510, 5929

\bibitem[{{Charisi} {et~al.}(2024){Charisi}, {Taylor}, {Witt}, \& {Runnoe}}]{2024PhRvL.132f1401C}
{Charisi}, M., {Taylor}, S.~R., {Witt}, C.~A., \& {Runnoe}, J. 2024, \prl, 132, 061401

\bibitem[{Cornish \& Sesana(2013)}]{Cornish_2013}
Cornish, N.~J. \& Sesana, A. 2013, Classical and Quantum Gravity, 30, 224005

\bibitem[{Damour \& Vilenkin(2000)}]{Damour_2000}
Damour, T. \& Vilenkin, A. 2000, Physical Review Letters, 85, 3761–3764

\bibitem[{Ellis \& van Haasteren(2017)}]{justin_ellis_2017_1037579}
Ellis, J. \& van Haasteren, R. 2017, jellis18/PTMCMCSampler: Official Release

\bibitem[{Ellis(2013)}]{Ellis_2013}
Ellis, J.~A. 2013, Classical and Quantum Gravity, 30, 224004

\bibitem[{{Ellis} {et~al.}(2012){Ellis}, {Siemens}, \& {Creighton}}]{2012ApJ...756..175E}
{Ellis}, J.~A., {Siemens}, X., \& {Creighton}, J.~D.~E. 2012, \apj, 756, 175

\bibitem[{{Ellis} {et~al.}(2019){Ellis}, {Vallisneri}, {Taylor}, \& {Baker}}]{enterprise}
{Ellis}, J.~A., {Vallisneri}, M., {Taylor}, S.~R., \& {Baker}, P.~T. 2019, {ENTERPRISE: Enhanced Numerical Toolbox Enabling a Robust PulsaR Inference SuitE}, Astrophysics Source Code Library, record ascl:1912.015

\bibitem[{{EPTA Collaboration} {et~al.}(2023{\natexlab{a}}){EPTA Collaboration}, {Antoniadis}, {Babak}, {Bak Nielsen}, {Bassa}, {Berthereau}, {Bonetti}, {Bortolas}, {Brook}, {Burgay}, {Caballero}, {Chalumeau}, {Champion}, {Chanlaridis}, {Chen}, {Cognard}, {Desvignes}, {Falxa}, {Ferdman}, {Franchini}, {Gair}, {Goncharov}, {Graikou}, {Grie{\ss}meier}, {Guillemot}, {Guo}, {Hu}, {Iraci}, {Izquierdo-Villalba}, {Jang}, {Jawor}, {Janssen}, {Jessner}, {Karuppusamy}, {Keane}, {Keith}, {Kramer}, {Krishnakumar}, {Lackeos}, {Lee}, {Liu}, {Liu}, {Lyne}, {McKee}, {Main}, {Mickaliger}, {Ni{\c{t}}u}, {Parthasarathy}, {Perera}, {Perrodin}, {Petiteau}, {Porayko}, {Possenti}, {Quelquejay Leclere}, {Samajdar}, {Sanidas}, {Sesana}, {Shaifullah}, {Speri}, {Spiewak}, {Stappers}, {Susarla}, {Theureau}, {Tiburzi}, {van der Wateren}, {Vecchio}, {Venkatraman Krishnan}, {Verbiest}, {Wang}, {Wang}, \& {Wu}}]{2023A&A...678A..48E}
{EPTA Collaboration}, {Antoniadis}, J., {Babak}, S., {et~al.} 2023{\natexlab{a}}, \aap, 678, A48

\bibitem[{{EPTA Collaboration} {et~al.}(2023{\natexlab{b}}){EPTA Collaboration}, {InPTA Collaboration}, {Antoniadis}, {Arumugam}, {Arumugam}, {Babak}, {Bagchi}, {Bak Nielsen}, {Bassa}, {Bathula}, {Berthereau}, {Bonetti}, {Bortolas}, {Brook}, {Burgay}, {Caballero}, {Chalumeau}, {Champion}, {Chanlaridis}, {Chen}, {Cognard}, {Dandapat}, {Deb}, {Desai}, {Desvignes}, {Dhanda-Batra}, {Dwivedi}, {Falxa}, {Ferdman}, {Franchini}, {Gair}, {Goncharov}, {Gopakumar}, {Graikou}, {Grie{\ss}meier}, {Guillemot}, {Guo}, {Gupta}, {Hisano}, {Hu}, {Iraci}, {Izquierdo-Villalba}, {Jang}, {Jawor}, {Janssen}, {Jessner}, {Joshi}, {Kareem}, {Karuppusamy}, {Keane}, {Keith}, {Kharbanda}, {Kikunaga}, {Kolhe}, {Kramer}, {Krishnakumar}, {Lackeos}, {Lee}, {Liu}, {Liu}, {Lyne}, {McKee}, {Maan}, {Main}, {Mickaliger}, {Ni{\c{t}}u}, {Nobleson}, {Paladi}, {Parthasarathy}, {Perera}, {Perrodin}, {Petiteau}, {Porayko}, {Possenti}, {Prabu}, {Quelquejay Leclere}, {Rana}, {Samajdar}, {Sanidas}, {Sesana}, {Shaifullah}, {Singha}, {Speri}, {Spiewak},
  {Srivastava}, {Stappers}, {Surnis}, {Susarla}, {Susobhanan}, {Takahashi}, {Tarafdar}, {Theureau}, {Tiburzi}, {van der Wateren}, {Vecchio}, {Venkatraman Krishnan}, {Verbiest}, {Wang}, {Wang}, \& {Wu}}]{EPTA}
{EPTA Collaboration}, {InPTA Collaboration}, {Antoniadis}, J., {et~al.} 2023{\natexlab{b}}, \aap, 678, A50

\bibitem[{{EPTA Collaboration} {et~al.}(2024){EPTA Collaboration}, {InPTA Collaboration}, {Antoniadis}, {Arumugam}, {Arumugam}, {Babak}, {Bagchi}, {Bak Nielsen}, {Bassa}, {Bathula}, {Berthereau}, {Bonetti}, {Bortolas}, {Brook}, {Burgay}, {Caballero}, {Chalumeau}, {Champion}, {Chanlaridis}, {Chen}, {Cognard}, {Dandapat}, {Deb}, {Desai}, {Desvignes}, {Dhanda-Batra}, {Dwivedi}, {Falxa}, {Ferdman}, {Franchini}, {Gair}, {Goncharov}, {Gopakumar}, {Graikou}, {Grie{\ss}meier}, {Gualandris}, {Guillemot}, {Guo}, {Gupta}, {Hisano}, {Hu}, {Iraci}, {Izquierdo-Villalba}, {Jang}, {Jawor}, {Janssen}, {Jessner}, {Joshi}, {Kareem}, {Karuppusamy}, {Keane}, {Keith}, {Kharbanda}, {Kikunaga}, {Kolhe}, {Kramer}, {Krishnakumar}, {Lackeos}, {Lee}, {Liu}, {Liu}, {Lyne}, {McKee}, {Maan}, {Main}, {Mickaliger}, {Ni{\c{t}}u}, {Nobleson}, {Paladi}, {Parthasarathy}, {Perera}, {Perrodin}, {Petiteau}, {Porayko}, {Possenti}, {Prabu}, {Quelquejay Leclere}, {Rana}, {Samajdar}, {Sanidas}, {Sesana}, {Shaifullah}, {Singha}, {Speri}, {Spiewak},
  {Srivastava}, {Stappers}, {Surnis}, {Susarla}, {Susobhanan}, {Takahashi}, {Tarafdar}, {Theureau}, {Tiburzi}, {van der Wateren}, {Vecchio}, {Venkatraman Krishnan}, {Verbiest}, {Wang}, {Wang}, {Wu}, {Auclair}, {Barausse}, {Caprini}, {Crisostomi}, {Fastidio}, {Khizriev}, {Middleton}, {Neronov}, {Postnov}, {Roper Pol}, {Semikoz}, {Smarra}, {Steer}, {Truant}, \& {Valtolina}}]{2024A&A...685A..94E}
{EPTA Collaboration}, {InPTA Collaboration}, {Antoniadis}, J., {et~al.} 2024, \aap, 685, A94

\bibitem[{{EPTA Collaboration} {et~al.}(2023{\natexlab{c}}){EPTA Collaboration}, {InPTA Collaboration}, {Antoniadis}, {Arumugam}, {Arumugam}, {Babak}, {Bagchi}, {Nielsen}, {Bassa}, {Bathula}, {Berthereau}, {Bonetti}, {Bortolas}, {Brook}, {Burgay}, {Caballero}, {Chalumeau}, {Champion}, {Chanlaridis}, {Chen}, {Cognard}, {Dandapat}, {Deb}, {Desai}, {Desvignes}, {Dhanda-Batra}, {Dwivedi}, {Falxa}, {Ferdman}, {Franchini}, {Gair}, {Goncharov}, {Gopakumar}, {Graikou}, {Grie{\ss}meier}, {Guillemot}, {Guo}, {Gupta}, {Hisano}, {Hu}, {Iraci}, {Izquierdo-Villalba}, {Jang}, {Jawor}, {Janssen}, {Jessner}, {Joshi}, {Kareem}, {Karuppusamy}, {Keane}, {Keith}, {Kharbanda}, {Kikunaga}, {Kolhe}, {Kramer}, {Krishnakumar}, {Lackeos}, {Lee}, {Liu}, {Liu}, {Lyne}, {McKee}, {Maan}, {Main}, {Mickaliger}, {Ni{\c{t}}u}, {Nobleson}, {Paladi}, {Parthasarathy}, {Perera}, {Perrodin}, {Petiteau}, {Porayko}, {Possenti}, {Prabu}, {Leclere}, {Rana}, {Samajdar}, {Sanidas}, {Sesana}, {Shaifullah}, {Singha}, {Speri}, {Spiewak}, {Srivastava},
  {Stappers}, {Surnis}, {Susarla}, {Susobhanan}, {Takahashi}, {Tarafdar}, {Theureau}, {Tiburzi}, {van der Wateren}, {Vecchio}, {Krishnan}, {Verbiest}, {Wang}, {Wang}, \& {Wu}}]{2023}
{EPTA Collaboration}, {InPTA Collaboration}, {Antoniadis}, J., {et~al.} 2023{\natexlab{c}}, \aap, 678, A49

\bibitem[{{Freedman} \& {Vigeland}(2024)}]{2024arXiv240711135F}
{Freedman}, G.~E. \& {Vigeland}, S.~J. 2024, arXiv e-prints, arXiv:2407.11135

\bibitem[{{Gair} {et~al.}(2014){Gair}, {Romano}, {Taylor}, \& {Mingarelli}}]{2014PhRvD..90h2001G}
{Gair}, J., {Romano}, J.~D., {Taylor}, S., \& {Mingarelli}, C. M.~F. 2014, \prd, 90, 082001

\bibitem[{{Gardiner} {et~al.}(2024){Gardiner}, {Kelley}, {Lemke}, \& {Mitridate}}]{2024ApJ...965..164G}
{Gardiner}, E.~C., {Kelley}, L.~Z., {Lemke}, A.-M., \& {Mitridate}, A. 2024, \apj, 965, 164

\bibitem[{{Goldstein} {et~al.}(2019){Goldstein}, {Sesana}, {Holgado}, \& {Veitch}}]{2019MNRAS.485..248G}
{Goldstein}, J.~M., {Sesana}, A., {Holgado}, A.~M., \& {Veitch}, J. 2019, \mnras, 485, 248

\bibitem[{{Goldstein} {et~al.}(2018){Goldstein}, {Veitch}, {Sesana}, \& {Vecchio}}]{2018MNRAS.477.5447G}
{Goldstein}, J.~M., {Veitch}, J., {Sesana}, A., \& {Vecchio}, A. 2018, \mnras, 477, 5447

\bibitem[{Hee {et~al.}(2015)Hee, Handley, Hobson, \& Lasenby}]{10.1093/mnras/stv2217}
Hee, S., Handley, W.~J., Hobson, M.~P., \& Lasenby, A.~N. 2015, Monthly Notices of the Royal Astronomical Society, 455, 2461

\bibitem[{{Hellings} \& {Downs}(1983)}]{1983_HD}
{Hellings}, R.~W. \& {Downs}, G.~S. 1983, \apjl, 265, L39

\bibitem[{Hourihane {et~al.}(2023)Hourihane, Meyers, Johnson, Chatziioannou, \& Vallisneri}]{Hourihane_2023}
Hourihane, S., Meyers, P., Johnson, A., Chatziioannou, K., \& Vallisneri, M. 2023, Physical Review D, 107

\bibitem[{{Jaffe} \& {Backer}(2003)}]{Jaffe2003}
{Jaffe}, A.~H. \& {Backer}, D.~C. 2003, \apj, 583, 616

\bibitem[{{Kelley} {et~al.}(2019){Kelley}, {Charisi}, {Burke-Spolaor}, {Simon}, {Blecha}, {Bogdanovic}, {Colpi}, {Comerford}, {D'Orazio}, {Dotti}, {Eracleous}, {Graham}, {Greene}, {Haiman}, {Holley-Bockelmann}, {Kara}, {Kelly}, {Komossa}, {Larson}, {Liu}, {Ma}, {Noble}, {Paschalidis}, {Rafikov}, {Ravi}, {Runnoe}, {Sesana}, {Stern}, {Strauss}, {U}, {Volonteri}, \& {Nanograv Collaboration}}]{2019BAAS...51c.490K}
{Kelley}, L., {Charisi}, M., {Burke-Spolaor}, S., {et~al.} 2019, \baas, 51, 490

\bibitem[{{Kocsis} \& {Sesana}(2011)}]{2011MNRAS.411.1467K}
{Kocsis}, B. \& {Sesana}, A. 2011, \mnras, 411, 1467

\bibitem[{{Kosowsky} {et~al.}(1992){Kosowsky}, {Turner}, \& {Watkins}}]{1992PhRvL..69.2026K}
{Kosowsky}, A., {Turner}, M.~S., \& {Watkins}, R. 1992, \prl, 69, 2026

\bibitem[{{Lentati} {et~al.}(2014){Lentati}, {Alexander}, {Hobson}, {Feroz}, {van Haasteren}, {Lee}, \& {Shannon}}]{2014MNRAS.437.3004L}
{Lentati}, L., {Alexander}, P., {Hobson}, M.~P., {et~al.} 2014, \mnras, 437, 3004

\bibitem[{{Lentati} {et~al.}(2015){Lentati}, {Taylor}, {Mingarelli}, {Sesana}, {Sanidas}, {Vecchio}, {Caballero}, {Lee}, {van Haasteren}, {Babak}, {Bassa}, {Brem}, {Burgay}, {Champion}, {Cognard}, {Desvignes}, {Gair}, {Guillemot}, {Hessels}, {Janssen}, {Karuppusamy}, {Kramer}, {Lassus}, {Lazarus}, {Liu}, {Os{\l}owski}, {Perrodin}, {Petiteau}, {Possenti}, {Purver}, {Rosado}, {Smits}, {Stappers}, {Theureau}, {Tiburzi}, \& {Verbiest}}]{2015MNRAS.453.2576L}
{Lentati}, L., {Taylor}, S.~R., {Mingarelli}, C.~M.~F., {et~al.} 2015, \mnras, 453, 2576

\bibitem[{{Liu} \& {Vigeland}(2021)}]{2021ApJ...921..178L}
{Liu}, T. \& {Vigeland}, S.~J. 2021, \apj, 921, 178

\bibitem[{{Petrov} {et~al.}(2024){Petrov}, {Taylor}, {Charisi}, \& {Ma}}]{2024arXiv240604409P}
{Petrov}, P., {Taylor}, S.~R., {Charisi}, M., \& {Ma}, C.-P. 2024, arXiv e-prints, arXiv:2406.04409

\bibitem[{{Rajagopal} \& {Romani}(1995)}]{1995ApJ...446..543R}
{Rajagopal}, M. \& {Romani}, R.~W. 1995, \apj, 446, 543

\bibitem[{{Ravi} {et~al.}(2014){Ravi}, {Wyithe}, {Shannon}, {Hobbs}, \& {Manchester}}]{2014MNRAS.442...56R}
{Ravi}, V., {Wyithe}, J.~S.~B., {Shannon}, R.~M., {Hobbs}, G., \& {Manchester}, R.~N. 2014, \mnras, 442, 56

\bibitem[{Reardon {et~al.}(2023)Reardon, Zic, Shannon, Hobbs, Bailes, Marco, Kapur, Rogers, Thrane, Askew, Bhat, Cameron, Cury{\l}o, Coles, Dai, Goncharov, Kerr, Kulkarni, Levin, Lower, Manchester, Mandow, Miles, Nathan, Os{\l}owski, Russell, Spiewak, Zhang, \& Zhu}]{PPTA}
Reardon, D.~J., Zic, A., Shannon, R.~M., {et~al.} 2023, \apjl, 951, L6

\bibitem[{Rosado {et~al.}(2015)Rosado, Sesana, \& Gair}]{Rosado_2015}
Rosado, P.~A., Sesana, A., \& Gair, J. 2015, Monthly Notices of the Royal Astronomical Society, 451, 2417–2433

\bibitem[{{Sah} {et~al.}(2024){Sah}, {Mukherjee}, {Saeedzadeh}, {Babul}, {Tremmel}, \& {Quinn}}]{2024arXiv240414508S}
{Sah}, M.~R., {Mukherjee}, S., {Saeedzadeh}, V., {et~al.} 2024, arXiv e-prints, arXiv:2404.14508

\bibitem[{Sesana(2013)}]{Sesana_2013}
Sesana, A. 2013, Classical and Quantum Gravity, 30, 244009

\bibitem[{{Sesana}(2013)}]{2013CQGra..30v4014S}
{Sesana}, A. 2013, Classical and Quantum Gravity, 30, 224014

\bibitem[{{Sesana} {et~al.}(2012){Sesana}, {Roedig}, {Reynolds}, \& {Dotti}}]{2012MNRAS.420..860S}
{Sesana}, A., {Roedig}, C., {Reynolds}, M.~T., \& {Dotti}, M. 2012, \mnras, 420, 860

\bibitem[{{Sesana} \& {Vecchio}(2010)}]{2010PhRvD..81j4008S}
{Sesana}, A. \& {Vecchio}, A. 2010, \prd, 81, 104008

\bibitem[{{Sesana} {et~al.}(2008){Sesana}, {Vecchio}, \& {Colacino}}]{2008MNRAS.390..192S}
{Sesana}, A., {Vecchio}, A., \& {Colacino}, C.~N. 2008, \mnras, 390, 192

\bibitem[{Taylor {et~al.}(2021)Taylor, Baker, Hazboun, Simon, \& Vigeland}]{enterprise_extension}
Taylor, S.~R., Baker, P.~T., Hazboun, J.~S., Simon, J., \& Vigeland, S.~J. 2021, enterprise\_extensions, v2.4.3

\bibitem[{{Taylor} \& {Gair}(2013)}]{2013PhRvD..88h4001T}
{Taylor}, S.~R. \& {Gair}, J.~R. 2013, \prd, 88, 084001

\bibitem[{{Taylor} {et~al.}(2020){Taylor}, {van Haasteren}, \& {Sesana}}]{2020PhRvD.102h4039T}
{Taylor}, S.~R., {van Haasteren}, R., \& {Sesana}, A. 2020, \prd, 102, 084039

\bibitem[{Taylor {et~al.}(2020)Taylor, van Haasteren, \& Sesana}]{PhysRevD.102.084039}
Taylor, S.~R., van Haasteren, R., \& Sesana, A. 2020, Phys. Rev. D, 102, 084039

\bibitem[{{Tiburzi} {et~al.}(2016){Tiburzi}, {Hobbs}, {Kerr}, {Coles}, {Dai}, {Manchester}, {Possenti}, {Shannon}, \& {You}}]{2016MNRAS.455.4339T}
{Tiburzi}, C., {Hobbs}, G., {Kerr}, M., {et~al.} 2016, \mnras, 455, 4339

\bibitem[{{Tomita}(1967)}]{1967PThPh..37..831T}
{Tomita}, K. 1967, Progress of Theoretical Physics, 37, 831

\bibitem[{{Truant} {et~al.}(2024){Truant}, {Izquierdo-Villalba}, {Sesana}, {Mohiuddin Shaifullah}, \& {Bonetti}}]{2024arXiv240712078T}
{Truant}, R.~J., {Izquierdo-Villalba}, D., {Sesana}, A., {Mohiuddin Shaifullah}, G., \& {Bonetti}, M. 2024, arXiv e-prints, arXiv:2407.12078

\bibitem[{{Vallisneri}(2020)}]{2020ascl.soft02017V}
{Vallisneri}, M. 2020, {libstempo: Python wrapper for Tempo2}, Astrophysics Source Code Library, record ascl:2002.017

\bibitem[{{Vallisneri} {et~al.}(2024){Vallisneri}, {Crisostomi}, {Johnson}, \& {Meyers}}]{2024arXiv240508857V}
{Vallisneri}, M., {Crisostomi}, M., {Johnson}, A.~D., \& {Meyers}, P.~M. 2024, arXiv e-prints, arXiv:2405.08857

\bibitem[{{Valtolina} {et~al.}(2024){Valtolina}, {Shaifullah}, {Samajdar}, \& {Sesana}}]{2024A&A...683A.201V}
{Valtolina}, S., {Shaifullah}, G., {Samajdar}, A., \& {Sesana}, A. 2024, \aap, 683, A201

\bibitem[{Verbiest {et~al.}(2012)Verbiest, Weisberg, Chael, Lee, \& Lorimer}]{Verbiest_2012}
Verbiest, J. P.~W., Weisberg, J.~M., Chael, A.~A., Lee, K.~J., \& Lorimer, D.~R. 2012, The Astrophysical Journal, 755, 39

\bibitem[{{Wyithe} \& {Loeb}(2003)}]{2003ApJ...590..691W}
{Wyithe}, J. S.~B. \& {Loeb}, A. 2003, \apj, 590, 691

\bibitem[{{Xin} {et~al.}(2021){Xin}, {Mingarelli}, \& {Hazboun}}]{2021ApJ...915...97X}
{Xin}, C., {Mingarelli}, C. M.~F., \& {Hazboun}, J.~S. 2021, \apj, 915, 97

\bibitem[{Xu {et~al.}(2023)Xu, Chen, Guo, Jiang, Wang, Xu, Xue, Caballero, Yuan, Xu, Wang, Hao, Luo, Lee, Han, Jiang, Shen, Wang, Wang, Xu, Wu, Manchester, Qian, Guan, Huang, Sun, \& Zhu}]{CPTA}
Xu, H., Chen, S., Guo, Y., {et~al.} 2023, Research in Astronomy and Astrophysics, 23, 075024

\end{thebibliography}


\end{document}